\documentclass[12pt]{article}
\usepackage{epsfig,fullpage}

\newcommand{\One}{1\kern-4.5pt1}
\begin{document}
\begin{center}

\begin{flushright}
HU-EP-04/39\\
SFB/CPP-04-40
\end{flushright}
\par \vskip 10mm

\vskip 1.2in

\begin{center}
{\LARGE\bf Pion mass splitting and phase structure in Twisted Mass QCD. 
} 

\vskip 0.7in
L. Scorzato$\! ^a$ \\
\vskip 0.2in

$^a\,${\it Institut f\"ur Physik, Humboldt Universit\"at zu Berlin, 12489 - Germany,}
\end{center}

\vskip 1.0in 
{\large\bf Abstract.}
In the framework of Wilson Chiral Perturbation Theory \cite{Rupak:2002sm}, we study the effect
induced by a twisted Wilson term, as it appears in Twisted Mass QCD \cite{Frezzotti:2000nk}
(with 2 degenerate quarks). 
In particular we consider the vacuum orientation and the pion masses. 
The computations are done to NLO both in the mass and in the lattice spacing (i.e. to $O(a^2)$). 
There are no restrictions on the relative size of lattice artifacts with respect to the physical 
mass, thus allowing, in principle, to bridge between the physical regime and the unphysical 
one, where lattice artifacts tend to dominate. 
The inclusion of $O(a^2)$ lattice artifacts can account for the splitting of degeneracy of the
three pion masses. Moreover $O(a^2)$ terms are necessary to model non trivial behaviors of the 
vacuum orientation such as possible Aoki phases. It turns out that these last two phenomena
are determined by the same constant.
\end{center}

\newpage

\section{Introduction and Motivations}
Twisted mass QCD (tmQCD) \cite{Frezzotti:2000nk} is a promising regularization for Lattice QCD, 
because it might succeed in improving on many of the most annoying problems of Lattice
QCD with Wilson-type fermions. 
It should be able to provide an easy way to reduce lattice artifacts, 
allow simulations of lighter quark masses and even of non degenerate quark masses, still with a positive
determinant \cite{Frezzotti:2003ni,Frezzotti:2004wz}. 
This is accomplished, basically,
by introducing a new  degree of freedom in Wilson Lattice QCD: the chiral-flavor 
orientation $\vec{\omega}$ of the subtracted Wilson term\footnote{We 
use a standard notation: $a$ is the lattice spacing,
$r$ the Wilson parameter, $g$ the gauge coupling, $M_{cr}(g)$ the critical mass,  $\nabla$ 
($\nabla^*$, $\tilde\nabla$) is the forward (backward, symmetrized)
covariant derivative, $\tau_i$ the Pauli matrices.}:
$W_{cr} = -\frac{a r}{2} \nabla^* \cdot \nabla + M_{cr}(g)$.
The most general Lattice Action, that we refer to, is:
\begin{equation} \label{eq:Action}
S_{\vec{\omega}}[U,\bar{\psi},\psi;g;M_q] = S_g[U;g^2]+ a^4 \sum_{xy}
\bar{\psi}_x \, [ \gamma \cdot \tilde\nabla +  
e^{i \gamma_5 \vec{\omega} \cdot \vec{\tau}} W_{cr}  + M_q]_{xy} \, \psi_y.
\end{equation}

Wilson Chiral Perturbation Theory (WChPT) \cite{Rupak:2002sm,Bar:2003mh}
has proved to be a crucial tool for the analysis of lattice data
\cite{Farchioni:2003bx,Farchioni:2003nf,Farchioni:2004tv,Aoki:2003yv}. 
WChPT has already been extended to the tmQCD case in \cite{Munster:2003ba,Munster:2004dj}, 
including $O(a)$ terms.

There are good reasons to consider also $O(a^2)$ corrections. In fact, as proposed
in \cite{Frezzotti:2003ni,Frezzotti:2004wz}, the special choice of twisting
$\omega=\pi/2$ leads to automatic $O(a)$ improvement of most observables, and thus
the leading lattice artifacts are necessarily of $O(a^2)$. Even if $\omega\neq\pi/2$,
the three pions of ($N_f=2$) tmQCD are still degenerate to $O(a)$. However tmQCD
breaks explicitly the flavor symmetry, for which the pion mass splitting is an
important signal. Such splitting  appears only at $O(a^2)$. Finally the inclusion
of $O(a^2)$ terms in the Chiral Lagrangian is necessary in order to describe the 
phase diagram of Wilson Lattice QCD at very small masses \cite{Aoki:1996ft,Sharpe:1998xm}.
It is natural to ask whether this is crucial also in presence of twisting, and
what is the scenario in this case.

We have a problem, however, with Twisted Wilson ChPT. In order to fully profit of an expansion
to $O(a^2)$, we would like to treat the mass $m$ 
and the lattice spacing $a$ on the same level, without requiring $m >> a \Lambda^2$ 
($\Lambda$ will be defined later). But if we do that, we do not have anymore a fixed vacuum around which
we can expand\footnote{We adopt the convention of calling ``vacuum'' the saddle
point of the Chiral Lagrangian around which the perturbative expansion is performed.}. 
In fact when $m$ and $a$ point to different directions, the limit $m\rightarrow 0$ and
$a\rightarrow 0$ do not commute. In principle, of course, the continuum limit 
should be performed first, and the chiral limit afterwards. But the whole idea of WChPT
is to reproduce the practical situation, that occurs in Monte Carlo simulations,
where both quantities are never small enough. 
Therefore we want to keep, as much as possible, a general set up. 
As we will see in section \ref{sec:Vacuum}, this
is just a technical difficulty that can be easily overcome. 

The purpose of this work is to provide a support to lattice unquenched simulations of tmQCD
which have been already started \cite{Farchioni:2004us,Ilgenfritz:2003gw}. 
For a recent review of ChPT at finite $a$ in general, we refer to \cite{Bar:lat04}.
Other works have been presented very recently studying related 
problems \cite{Munster:2004am,Wu:lat04,Sharpe:2004ps,Aoki:2004ta}.
We hope that from so much activity a clearer picture can emerge.

In section \ref{sec:Lag} we write explicitly the Chiral Lagrangian that we use, 
and give some comments. 
In section \ref{sec:Vacuum} we compute the vacuum orientation, and study 
the special behavior near the critical region. 
In section \ref{sec:Masses} we compute the pion masses and in section \ref{sec:disc} we
give a final discussion.

\section{Chiral Effective Lagrangian}
\label{sec:Lag}
The low energy description for Lattice tmQCD that we use in the present work,
is given by the effective Lagrangian $\mathcal{L}_{\chi}= \mathcal{L}_{LO}+ \mathcal{L}_{NLO}$
\cite{Gasser:1984yg,Rupak:2002sm,Bar:2003mh,Bar:2003xq}:
\begin{eqnarray}\label{eq:ChLag}
\mathcal{L}_{LO}  &=& 
\frac{{F^2 }}{4}\left\langle \partial_{\mu} \Sigma\partial^{\mu} \Sigma^\dag
\right\rangle +
 \frac{{F^2 }}{4}\left\langle \hat m \Sigma^\dag + \Sigma
  \hat m\right\rangle +
\frac{{F^2 }}{4}\left\langle \hat a \Sigma^\dag+\Sigma \hat a^\dag \right\rangle,\\ 
\mathcal{L}_{NLO} &=& \;{L_1 \left\langle {\partial_{\mu} \Sigma\partial^{\mu} \Sigma^\dag
    } \right\rangle ^2 }+
{L_2 \left\langle {\partial_{\mu}  \Sigma\partial_\nu  \Sigma^\dag  }
  \right\rangle 
 \left\langle {\partial^\mu  \Sigma\partial^\nu  \Sigma^\dag  }
 \right\rangle }+
{L_3 \left\langle {(\partial_{\mu} \Sigma\partial^{\mu} \Sigma^\dag  )^2 }
  \right\rangle }+\nonumber\\ 
&&+\;{L_4 \left\langle {\partial_{\mu} \Sigma\partial^{\mu} \Sigma^\dag  }
 \right\rangle \left\langle 
{\hat m  \Sigma^\dag + \Sigma \hat m } \right\rangle }
+{W_4 \left\langle {\partial_{\mu} \Sigma\partial^{\mu} \Sigma^\dag  }
  \right\rangle \left\langle  
{\hat a \Sigma^\dag + \Sigma \hat a^\dag} \right\rangle }+\nonumber\\
&&+\;{L_5 \left\langle {\partial_{\mu} \Sigma\partial^{\mu} \Sigma^\dag 
 (\hat m  \Sigma^\dag + \Sigma \hat m )} \right\rangle }
+{W_5 \left\langle {\partial_{\mu} \Sigma\partial^{\mu} \Sigma^\dag  (  
\hat a \Sigma^\dag + \Sigma \hat a^\dag)} \right\rangle }+\nonumber\\
&&+\;{L_6 \left\langle {\hat m  \Sigma^\dag + \Sigma  \hat m }
  \right\rangle ^2 } 
+{W_{6} \left\langle {\hat m  \Sigma^\dag + \Sigma \hat m }
  \right\rangle \left\langle {\hat a \Sigma^\dag + \Sigma \hat a^\dag}
  \right\rangle }
+ {W'_6\left\langle \hat a \Sigma^\dag+\Sigma \hat a^\dag \right\rangle ^2}+
\nonumber\\ 
&&+\; {L_7 \left\langle {\hat m  \Sigma^\dag - \Sigma \hat m }
  \right\rangle ^2 } 
+{W_{7} \left\langle {\hat m  \Sigma^\dag - \Sigma \hat m
    } \right\rangle  
\left\langle {\hat a \Sigma^\dag - \Sigma \hat a^\dag} \right\rangle
}
+{W'_7\left\langle \hat a \Sigma^\dag-\Sigma \hat a^\dag \right\rangle^2}+
\nonumber\\ 
&&+\;{L_8 \left\langle {\hat m  \Sigma^\dag \hat m  \Sigma^\dag +
      \Sigma   
\hat m \Sigma \hat m } \right\rangle }
+{W_{8} \left\langle {\hat m \Sigma ^\dag \hat a \Sigma^\dag + \Sigma \hat a^\dag
 \Sigma \hat m } \right\rangle }
+{W'_8\left\langle \hat a \Sigma^\dag \hat a \Sigma^\dag+
\Sigma \hat a^\dag \Sigma \hat a^\dag \right\rangle}.\nonumber
\end{eqnarray}
The field $\Sigma(x)$ is related to the pion field by
\[
\Sigma(x) = e^{\frac{i \, \vec{\tau}\cdot \vec{\pi}(x)}{F}},
\]
where $F\simeq 93$MeV is the pion decay constant, 
and $\tau_j$ are the Pauli matrices. 
The coefficients $L_i$ \cite{Gasser:1984yg} are the usual dimensionless 
Gasser-Leutwyler's Low Energy Constants (LEC). 
The other dimensionless LEC's $W_i$ and the $W'_i$ -- 
introduced in \cite{Rupak:2002sm,Bar:2003mh} --  describe the effect of
the lattice artifacts which appear in the Wilson Lattice formulation of QCD (WLQCD) 
\cite{Wilson:1975id}.
Space time indices are summed according to the Minkowski metric (the Euclidean Lagrangian can 
be recovered by introducing a minus sign for each $(\partial \Sigma \partial \Sigma)$ term). 
In order to describe the tmQCD case, 
the parameters $\hat m$ and $\hat a$ are 2 by 2 matrices in flavor space. 
In this paper we always adopt the ``physical basis'' of 
tmQCD (see \cite{Frezzotti:2002iv}), in which case\footnote{
We also follow the choice of \cite{Bar:2003mh} where explicit reference to
$c_{SW}$ has been dropped.}
\begin{equation} \label{eq:m_a}
\hat m = 2 B_0
\left( 
\begin{array}{c c} 
m & 0 \\
0 & m
\end{array}
\right), \;\;\;\;\;\;\;
\hat a = 2 W_0
\left( 
\begin{array}{c c} 
a (\cos{\omega} + i \sin{\omega}) & 0 \\
0 & a (\cos{\omega} - i \sin{\omega}) 
\end{array}
\right). 
\end{equation} 
For later convenience we introduce the dimensionless parameters, 
(as already done in \cite{Farchioni:2003nf})
\[ 
\chi = \frac{2 B_0 m}{F^2}, \;\;\;\;\;\;\;\;\;\;\;\;
\rho = \frac{2 W_0 a}{F^2},\;\;\;\;\;\;\;\;\;\;\; 
\eta= \frac{\rho}{\chi}.
\]
We will never use the ``twisted basis'', in which lattice artifacts are $\propto 1_2$.
However the conversion is straightforward: 
$\mu_q = m \sin(\omega)$ and an ordinary mass $m_q= m \cos(\omega)$. 

The Lagrangian above represents an expansion for small momenta $p^2$, small quark
masses $m$ and small lattice spacing $a$. In particular we expect that it gives a good description
of WLQCD when $p^2$, $B_0 m$, $W_0 a$ $<< \Lambda^2 $, where $\Lambda \sim$ 1 GeV.
In (\ref{eq:ChLag}) all the terms necessary for a calculation to NLO of pion masses are included.
We refer to \cite{Rupak:2002sm,Bar:2003mh,Bar:2003xq}
for a full justification why no other terms need to be included at this order. 
As usual we choose $B_0>0$. In principle we do not know the sign of $W_0$,
but it can always be made positive by a redefinition of $\omega \rightarrow \omega + \pi$. Since 
we are going to study the whole range in $\omega$, the choice $W_0>0$ is not a real limitation.
For the same reason, we can restrict to $a>0$.
Finally the system at negative $m$ has a mirror description in the system
with positive $m$. Thus in the following we will assume $a,\; m,\; \eta >0$.

Here $m$ corresponds to ``some''  definition of the renormalized quark mass, and $a$ to
``some'' definition of the lattice spacing. By construction  of the Chiral Effective Lagrangian
the LEC's $L_i$, $W_i$, $W'_i$ do not depend on the {\em size} of $m$ or $a$, but they do depend
on the particular {\em definition} which is chosen for $m$ and $a$ (for instance based on pion
mass rather than PCAC)\footnote{
Of course the $W_i$ and $W'_i$ depend also on the choice of the lattice action.}.
In the continuum limit the physical $L_i$ are expected to loose also
this dependence, but we have no reason to think that the same will happen for the $W$'s. 
For instance if we redefine the mass and lattice spacing:
\begin{equation} \label{eq:matrans}
\hat m^* = \hat m + \hat a,     \;\;\;\;\;\;\;\;\; \hat a^* = \hat a,
\end{equation}
we find that the LEC's are correspondingly changed in
\begin{eqnarray} \label{eq:LECtrans}
L_{4,5,6,7,8}^* &=& L_{4,5,6,7,8}, \\
W_{4,5}^* &=& W_{4,5} - L_{4,5},  \nonumber\\
W_{6,7,8}^* &=& W_{6,7,8} - 2 L_{6,7,8}, \nonumber\\
{W'}_{6,7,8}^* &=& {W'}_{6,7,8} - W_{6,7,8} + L_{6,7,8},  \nonumber
\end{eqnarray}
and the LO $O(a)$ terms disappear, leaving only terms of order $O(p^2 \,a)$ and $O(m \,a)$.
We will occasionally employ $m^*$ later, which has both technical advantages (when 
computing logarithms) and provides an interesting point of view. But we mainly use 
the parameters $m$ and $a$, as in (\ref{eq:m_a}), because they are in the physical basis,
and have otherwise a generic form that represents your preferred definition of $m$ and $a$
(on the lattice $m^*$ is not always accessible).

To avoid confusion, let me state clearly that an $O(a)$ redefinition of the mass 
as in (\ref{eq:matrans}, \ref{eq:LECtrans}) does not
reshuffle the Symanzik expansion of course, which is well defined (given a lattice action),
since the mass term and the Pauli term are different operators. It is only
at the level of LO Chiral Lagrangian that the ambiguity appears, since at
that level we have essentially only one independent operator.

In order to make the contact with lattice simulations as clear as possible, let
me add a last remark 
(see \cite{Rupak:2002sm} for a slightly different formulation of the same concept).
The low energy representation (\ref{eq:ChLag})
of Lattice QCD is meant to hold for the renormalized Lagrangian. 
Changing the lattice spacing $a$ (typically through a change of $\beta$, 
or the Wilson parameter $r$) has big effects, at the level of the cutoff scale,
which cannot certainly be taken into account by the expansion in (\ref{eq:ChLag}). 
However this possibly strong dependence is compensated by the dependence on $a$
in the renormalization constants, 
delivering a renormalized action whose ``residual''  $a$ dependence is
parameterized in (\ref{eq:ChLag}). 

\section{Vacuum orientation}
\label{sec:Vacuum}
The first problem, that we have to consider, is the 
determination of the minimum of the potential derived from 
(\ref{eq:ChLag}) when the field $\Sigma(x)$ is set to a constant 2 by 2 unitary matrix
$\Sigma_0$ and the derivatives are set to zero.
In a typical ChPT calculation the vacuum orientation is changed by the introduction of a source.
However such sources are always extrapolated to zero in the final result, because they are needed
only to derive observables.
Even non degenerate quark masses do not produce a rotation of the vacuum in the chiral limit, 
since vector-flavor symmetry is not spontaneously broken. As a result $\Sigma_0=1$ in the final 
expressions. 
The case of ChPT for tmQCD is peculiar, because a vacuum rotation is produced by a non zero
$a$ and $\omega$, which are not sources, but true parameters of the model that we want to describe.
Moreover the vacuum orientation $\Sigma_0$ does not go to 1 even in the limit 
$a, m \rightarrow 0$.  In fact $\Sigma_0$ depends on the ratio $a/m$  in which the limit is attained.
This is obvious from the physical point of view, because the direction in which chiral symmetry
is spontaneously broken, depends on the direction, in chiral-flavor space, where the mass term is
pointing. At leading order in the effective theory the twisted lattice artifacts are nothing but
an effective mass term. In other words the limit $a\rightarrow 0$ and $m\rightarrow 0$ 
do not commute, and a double expansion in $a$ and $m$ cannot be defined around a single vacuum. 

Of course in the physical regime the continuum limit $a\rightarrow 0$ should always be taken before
the chiral limit $m\rightarrow 0$. Here the assumption $\eta << 1$ makes sense, and 
the point of minimum $\Sigma_0$  deviates from 1 only by small corrections of $O(a)$.
This is the approach adopted in \cite{Munster:2003ba,Munster:2004dj}.  
However, in this paper, we want to study also the regime where lattice artifacts 
are not necessarily  much smaller than the mass term. This is because we want to represent
the practical situation that occurs in Lattice Monte Carlo simulations, where the continuum
limit and the chiral limit cannot be completely and safely separated.  In fact, this is the whole
idea of WChPT.
Moreover we
also would like to describe the deep unphysical regime where lattice artifacts tend to
dominate over the mass term, which can have very interesting features, as first noticed
in \cite{Aoki:1996ft}.

The strategy of the present computation is 
to fix a ratio $\eta$ first, compute the corresponding vacuum $\Sigma_0(\eta)$, and
perform an expansion in $\chi$ for each fixed $\eta$. This is nothing but a convenient
organization of the computation, which let us read off, at the end, the results for different
regimes. Equivalently one could perform a ($\eta$-dependent)
change of variables to  $m^*$ in (\ref{eq:matrans}) and expand around the fixed 
vacuum determined by $m^*$ (which is however not fixed in the physical basis).

In order to find the minimum of the potential we use the parameterization:
\begin{equation} \label{eq:vacuum.param}
\Sigma_0 = 1_2 \cos(\theta) + i \tau_3 \sin(\theta) \cos(\phi) 
+ i \tau_1 \sin(\theta) \sin(\phi), 
\end{equation}
Having introduced a twisted term
in the $\tau_3$ direction, we expect a vacuum rotation in the same direction. However we cannot
-- a priori -- exclude a more complicated pattern. Since the directions $\tau_{1,2}$ are fully 
equivalent, the parameterization (\ref{eq:vacuum.param}) is completely general.

\subsection{Potential at LO}
\label{sec:LOvacuum}
\begin{figure}
\begin{center}
\includegraphics[height=5cm,width=5cm]{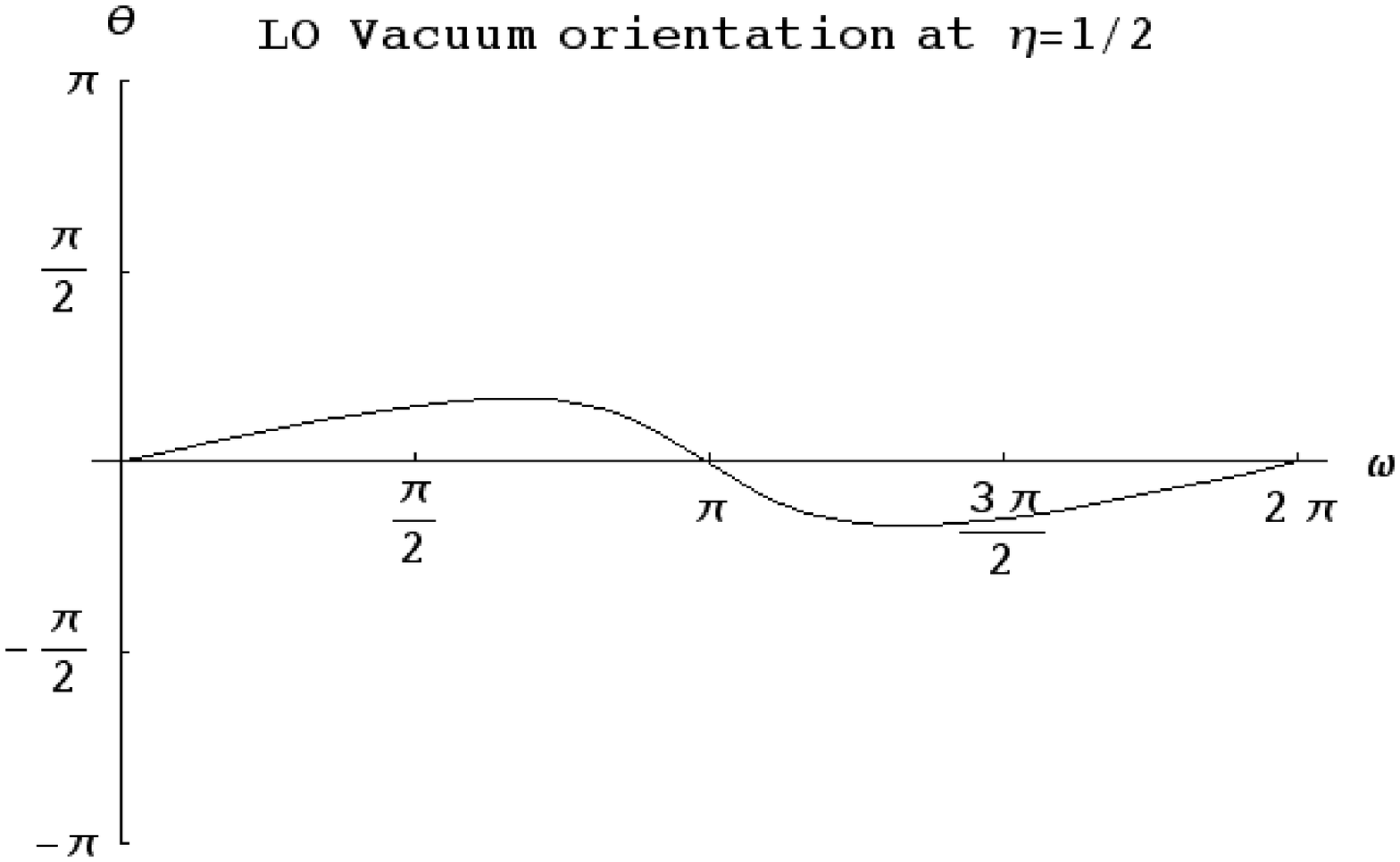}
\includegraphics[height=5cm,width=5cm]{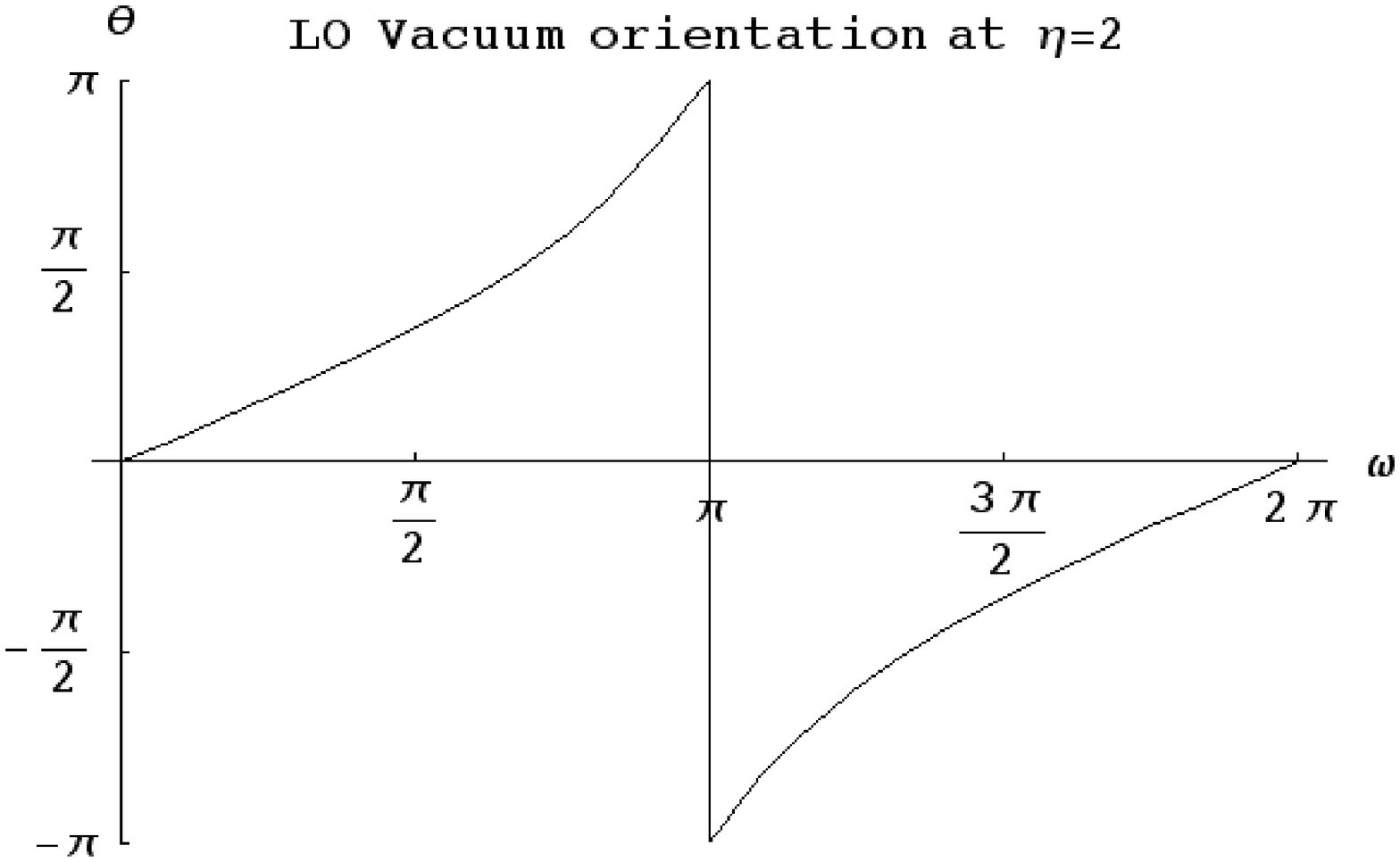}
\end{center}
\caption{\label{fig:LO}LO solution $\theta_{LO}$ as function of $\omega$.}
\end{figure}

At LO the potential is (proportional to): 
\begin{equation} \label{eq:LOpot}
{\cal V}_{\chi LO} =  
-\eta  \, \cos(\omega ) \, \cos(\theta )-\cos(\theta )-\eta  \, \cos(\phi ) \, 
\sin(\omega ) \, \sin(\theta )
\end{equation}
which has a minimum for $\phi=0$ and
\begin{equation} \label{eq:solOmegaLO}
\theta_{LO} \,  ={{\tan}^{-1}}
\bigg(\frac{\eta \,  \sin(\omega )}{\eta \,  \cos(\omega )+1}\bigg)\,  
+\,  \,  \Bigg( \begin{array}{ccc}
	  0  & \mbox{   if   } & {\;\; 1+ \eta \cos(\omega) > 0}\\
	 \pi & \mbox{   if   } & {\;\; 1+ \eta \cos(\omega) < 0 \; \&\& \; \sin(\omega) > 0}\\
	-\pi & \mbox{   if   } & {\;\; 1+ \eta \cos(\omega) < 0 \; \&\& \; \sin(\omega) < 0}
  \end{array} \Bigg)
\end{equation}
In our convention the Image of $\tan^{-1} = (-\pi/2,\pi/2)$.
The trivial orientation ($\theta=0$) is recovered whenever $\omega=0, 2 \pi$ or $\eta << 1$.
When  $\eta$ is not negligible, the vacuum undergoes a fluctuation as in figure \ref{fig:LO} (left),
but it is never rotated more than $\pi/2$, as long as $\eta<1$. When the lattice artifacts dominate
$\eta>1$, the vacuum undergoes a full rotation, tuned by $\omega$, 
as in figure \ref{fig:LO} (right). 
We see that for $\eta=1$ and $\omega=\pi$ the system is critical: it is the point where
the leading order ``effective mass'' $m^*$ changes sign. Here the LO terms in the potential cancel, 
and the NLO terms become relevant. Around this point one expects that phenomena like 
\cite{Aoki:1996ft,Sharpe:1998xm} may show up.
The fact that this happens at $\omega=\pi$ instead of $\omega=0$, is simply a convention related
to our choice $W_0>0$.

The stationary point (\ref{eq:solOmegaLO}) is indeed a minimum, since 
the second derivatives are always positive:
\begin{eqnarray} \label{eq:doublederiLO}
\frac{\partial^2}{\partial \theta^2} {\cal V}_{\chi}&=
{\sqrt{{{\eta }^2}+2  \cos(\omega )  \eta +1}} \\
\frac{\partial^2}{\partial \phi^2} {\cal V}_{\chi}&=
\frac{{{\eta }^2}  {{\sin}^2}(\omega )}
{{\sqrt{{{\eta }^2}+2  \cos(\omega )  \eta +1}}}\nonumber
\end{eqnarray}
Notice that 
$\frac{\partial}{\partial \phi}_{|\phi=0,\pi} {\cal V}_{\chi}(\theta)=0$ for each $\theta$
(this will be true also at NLO). 
Therefore the mixed second derivatives are always zero in $\phi=0,\pi$, which simplifies
considerably the study of the stability. Since only one maximum and one minimum
are possible in the function (\ref{eq:LOpot}), this conclude the study of the LO vacuum.

\subsection{Potential at NLO}
\label{sec:NLOvacuum}
If we add the NLO terms to the potential, we have:
\begin{eqnarray} 
{\cal V}_{\chi} &= {\cal V}_{\chi LO} - 
\chi [4\, \cos(2\, \theta )\, {L_{86}} + 
4\, (\cos(\omega )\, \cos(2\, \theta )+
\cos(\phi )\, \sin(\omega )\, \sin(2\, \theta ))\, {W_{86}}\, \eta \nonumber\\
&+
(
\cos(2 \theta )+ \cos(2 \omega ) +
3 \cos(2 \omega )  \cos(2 \theta )+ \nonumber\\
&+
4 \cos(2 \phi ) {{\sin}^2}(\omega ) {{\sin}^2}(\theta )+
4 \cos(\phi ) \sin(2 \omega ) \sin(2 \theta )
) {W'_{86}} {{\eta }^2} \label{eq:potNLO}
]
\end{eqnarray}
where we have defined 
$X_{86} = X_8 + 2 X_6$, for $X=L,W,W'$, which are the only relevant LEC's, here.

First of all we deal with the possibility of a non trivial stationary point in $\phi$. One finds
that, besides $\phi=0,\pi$, other possible solutions of 
$\frac{\partial}{\partial \phi} {\cal V}_{\chi}=0$ are:
\begin{equation} \label{eq:statTheta}
\phi = \pm   {{\cos}^{-1}}\bigg(-\frac{1+8  \chi   \cos(\theta )    {W_{86}}+16  \eta   \chi
  \cos(\omega )  \cos(\theta )  {{W'}_{86}}}{\sin(\theta )\sin(\omega )  16  \eta   \chi  {{W'}_{86}}}\bigg).
\end{equation}
However, in the regime where ChPT is applicable, the relation $\chi X << 1$ and $\chi \eta X << 1$
must hold for any LEC $X$. 
Therefore (\ref{eq:statTheta}) can never be a solution for our problem. It is known
\cite{Aoki:1996ft,Sharpe:1998xm} (and we will see it later in the context  of tmQCD), 
that NLO terms can actually produce new solutions for the minimum of the potential, but only
if LO terms are subject to cancellations, which are not possible in (\ref{eq:statTheta}). 
This seems to exclude the possibility of a phase of {\em spontaneously} broken flavor symmetry
(i.e. along a direction different from the one of twisting, when twisting is non zero), 
as it has been very recently suggested \cite{Munster:2004am}.
Although the general functional form of (\ref{eq:potNLO}),
would allow it, it is not within the reach of our ChPT representation.
This is an interesting question, which certainly deserve further numerical studies,
and may find a description within an alternative analytical approach.
However, in this paper, we set in the following $\phi=0$ 
(the choice $\phi=\pi$ is equivalent, if one correspondingly
changes $\theta \rightarrow - \theta$).

\begin{figure}
\includegraphics[height=5cm,width=5cm]{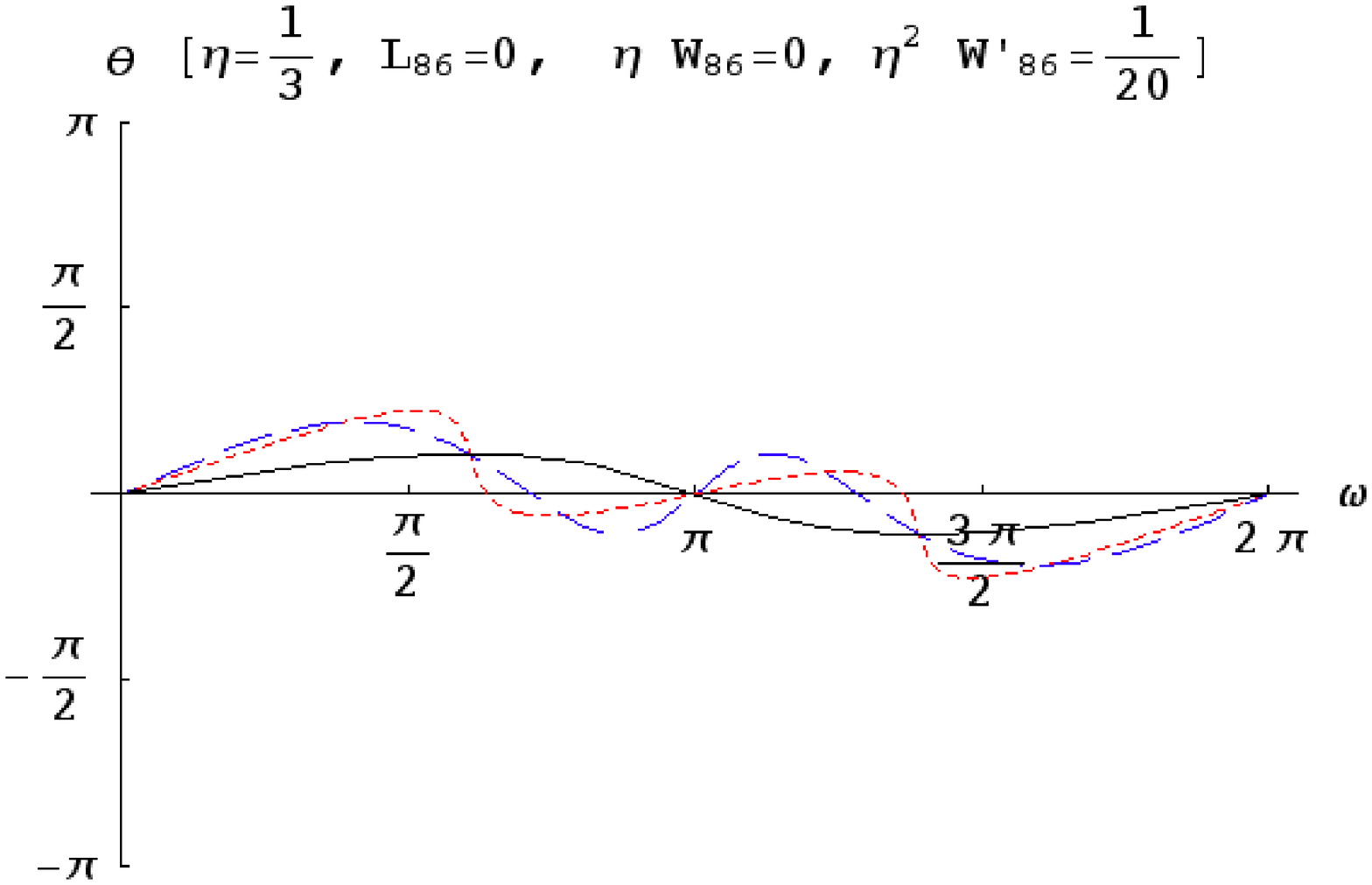}
\includegraphics[height=5cm,width=5cm]{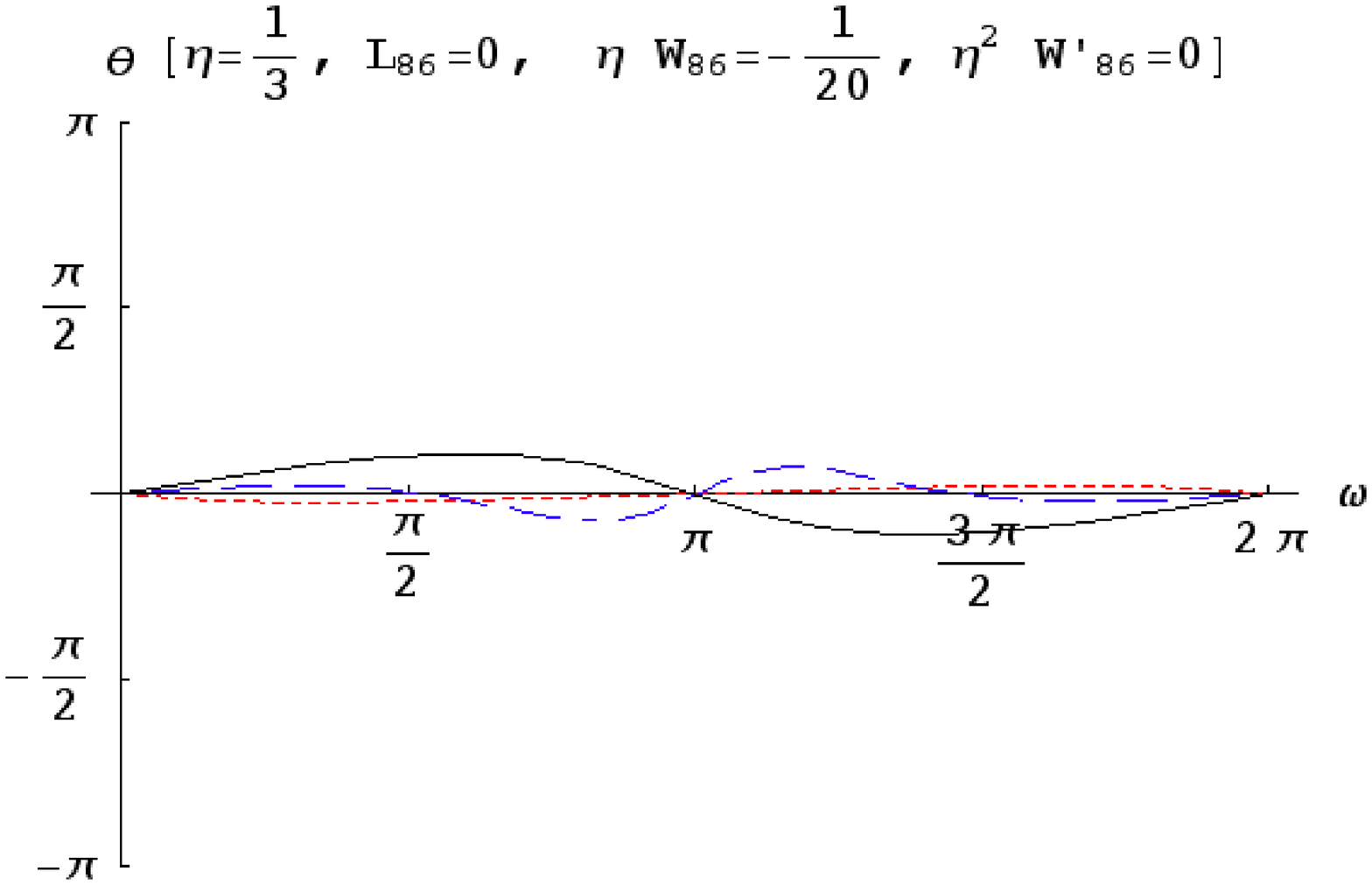}
\includegraphics[height=5cm,width=5cm]{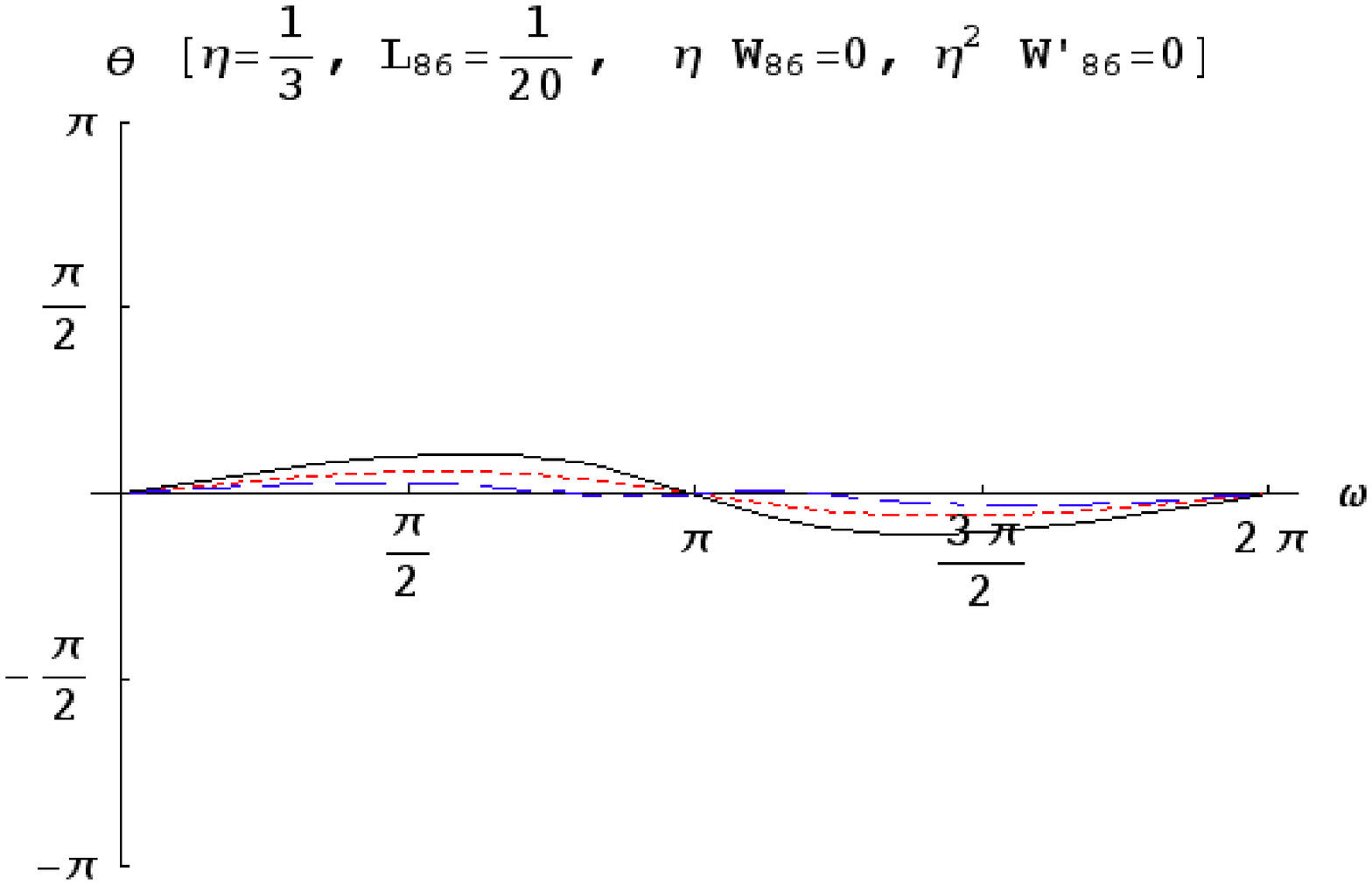}
\includegraphics[height=5cm,width=5cm]{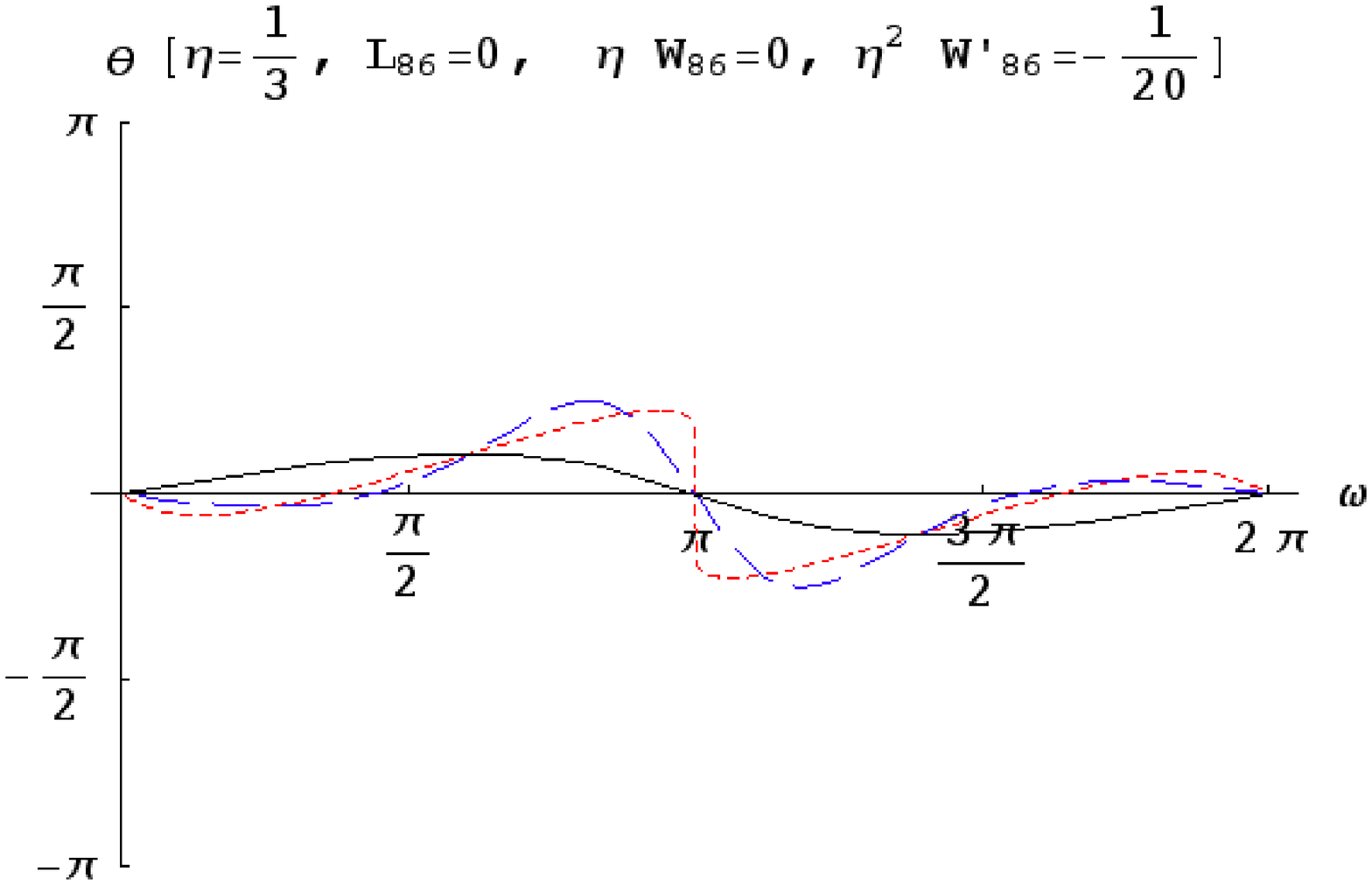}
\includegraphics[height=5cm,width=5cm]{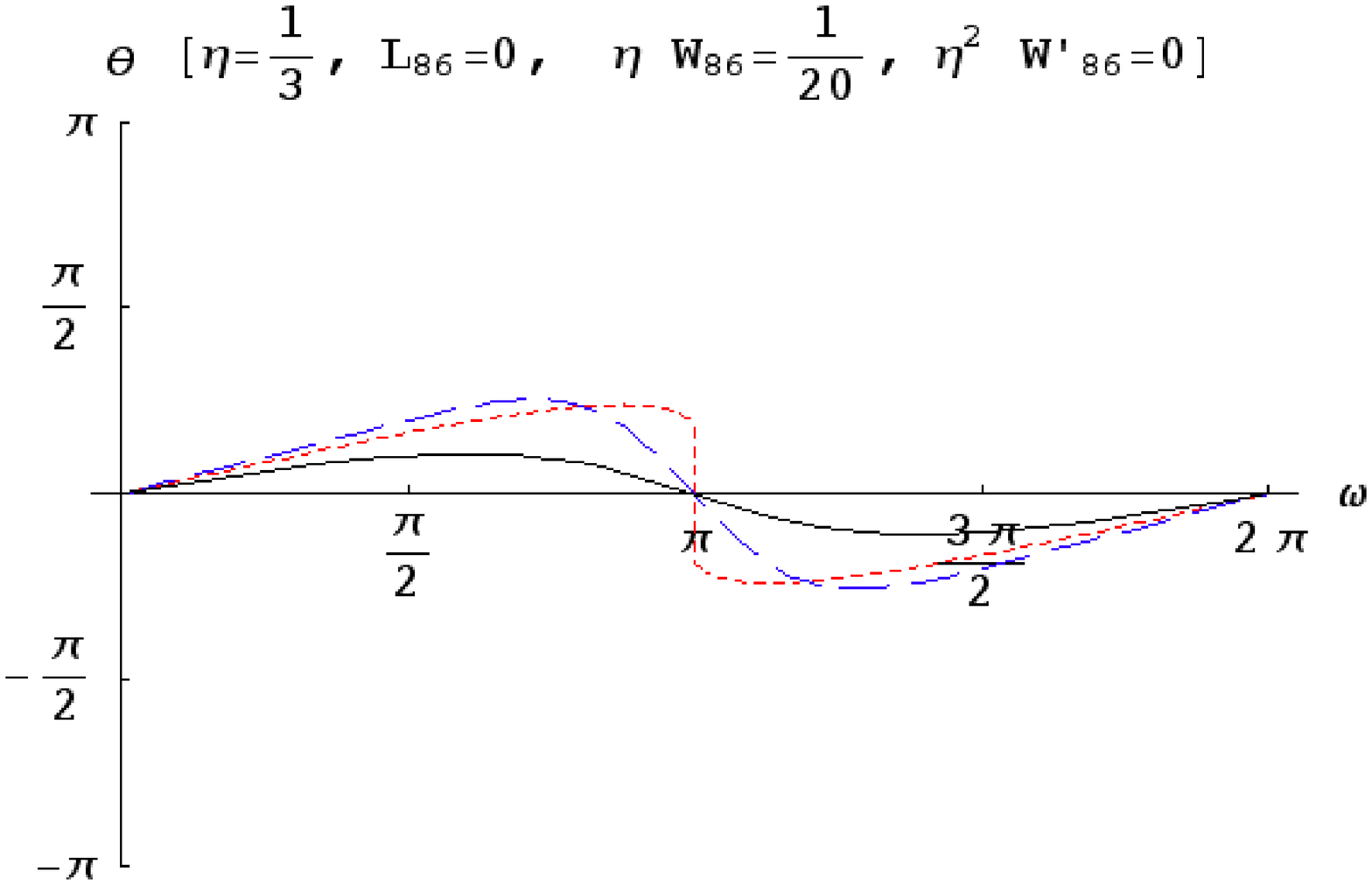}
\includegraphics[height=5cm,width=5cm]{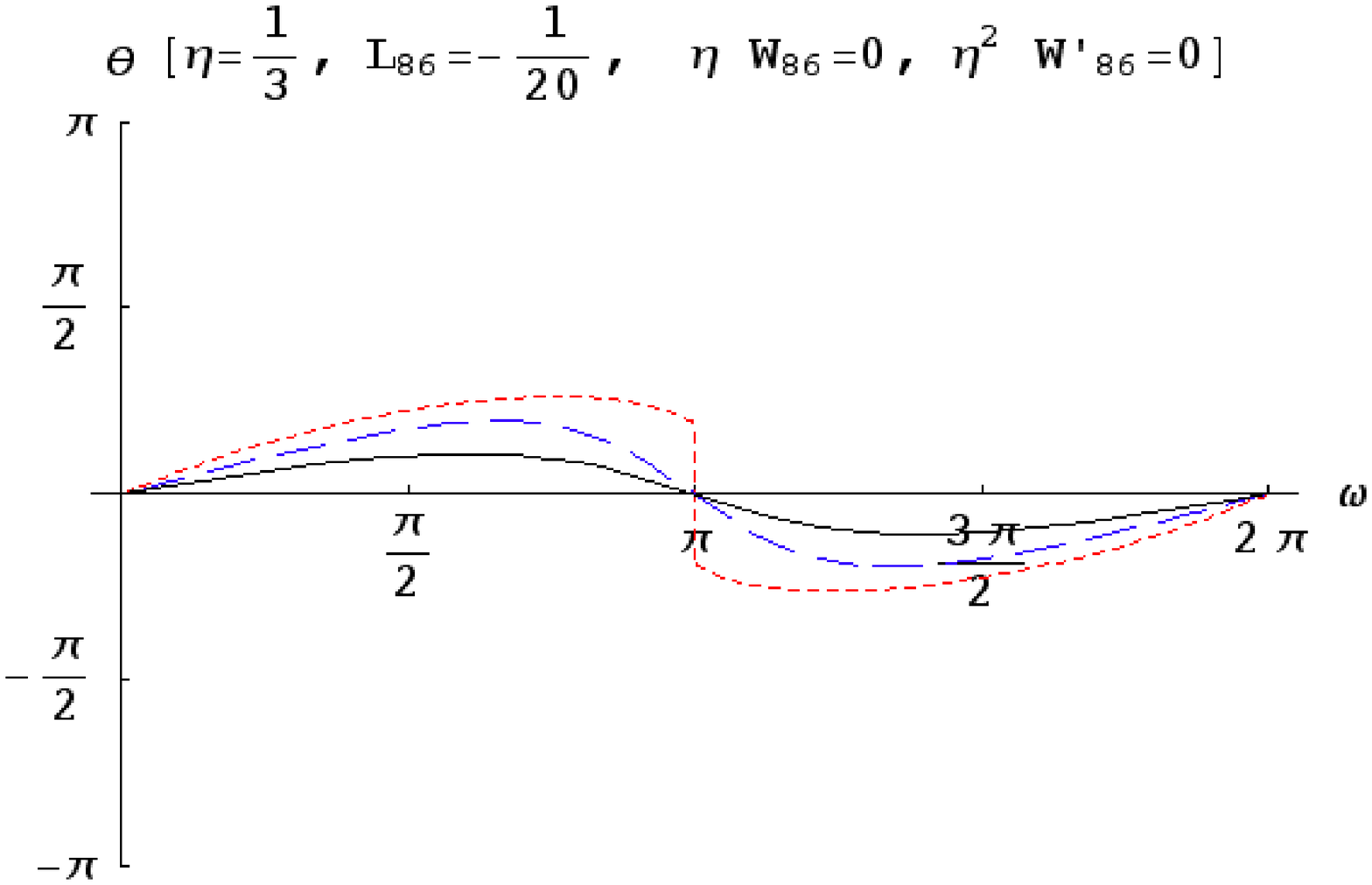}
\includegraphics[height=5cm,width=5cm]{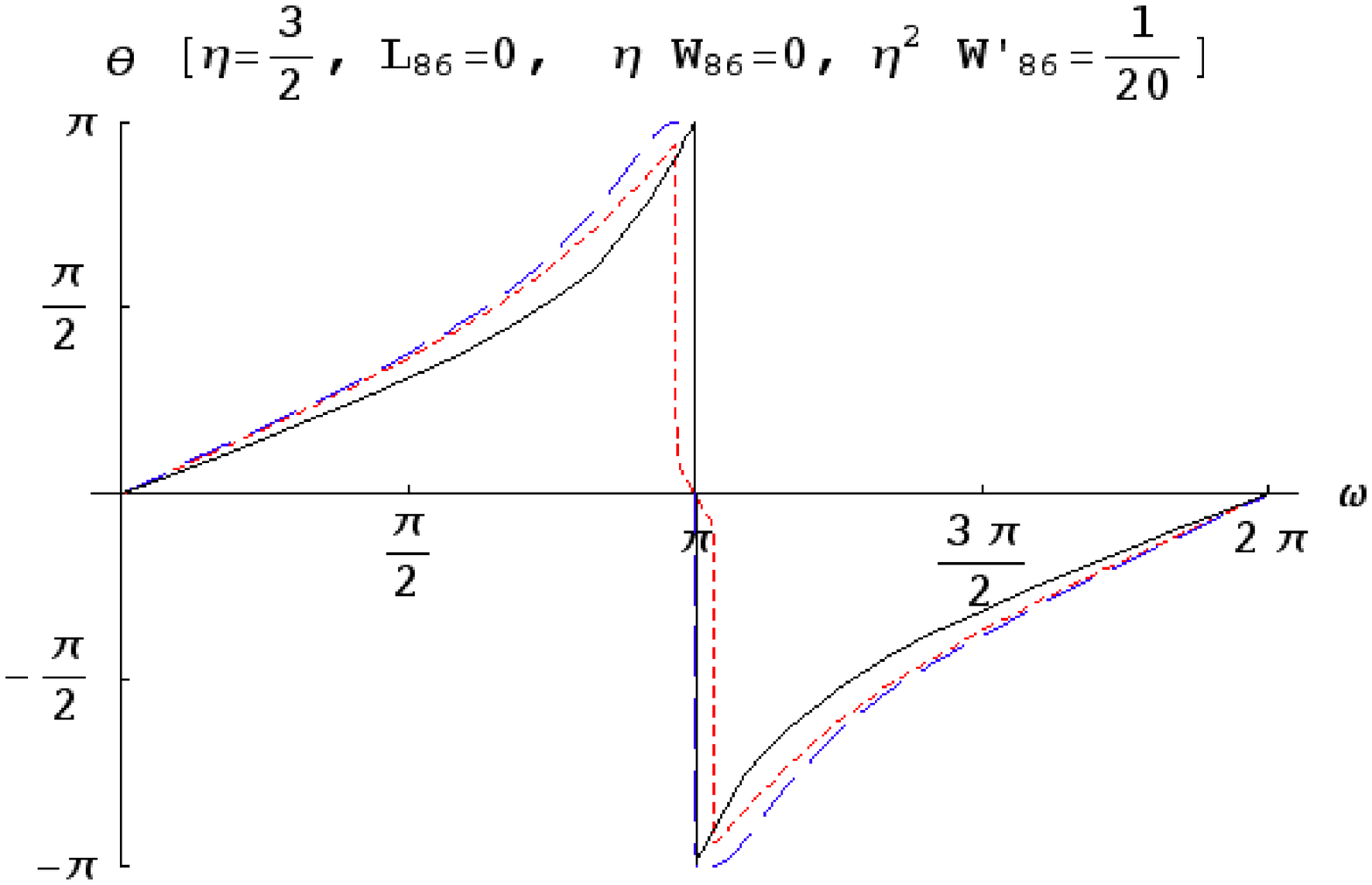}
\includegraphics[height=5cm,width=5cm]{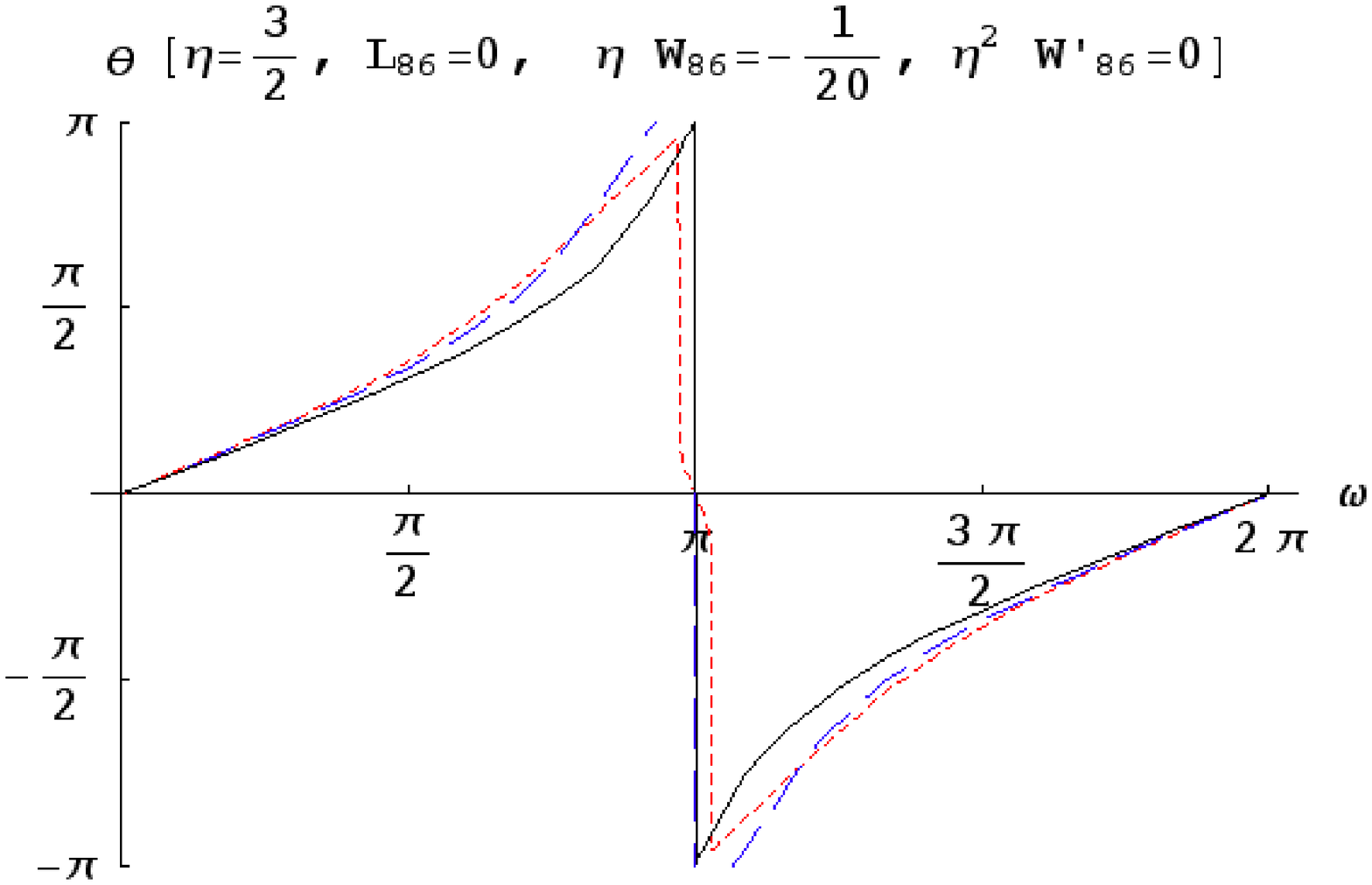}
\includegraphics[height=5cm,width=5cm]{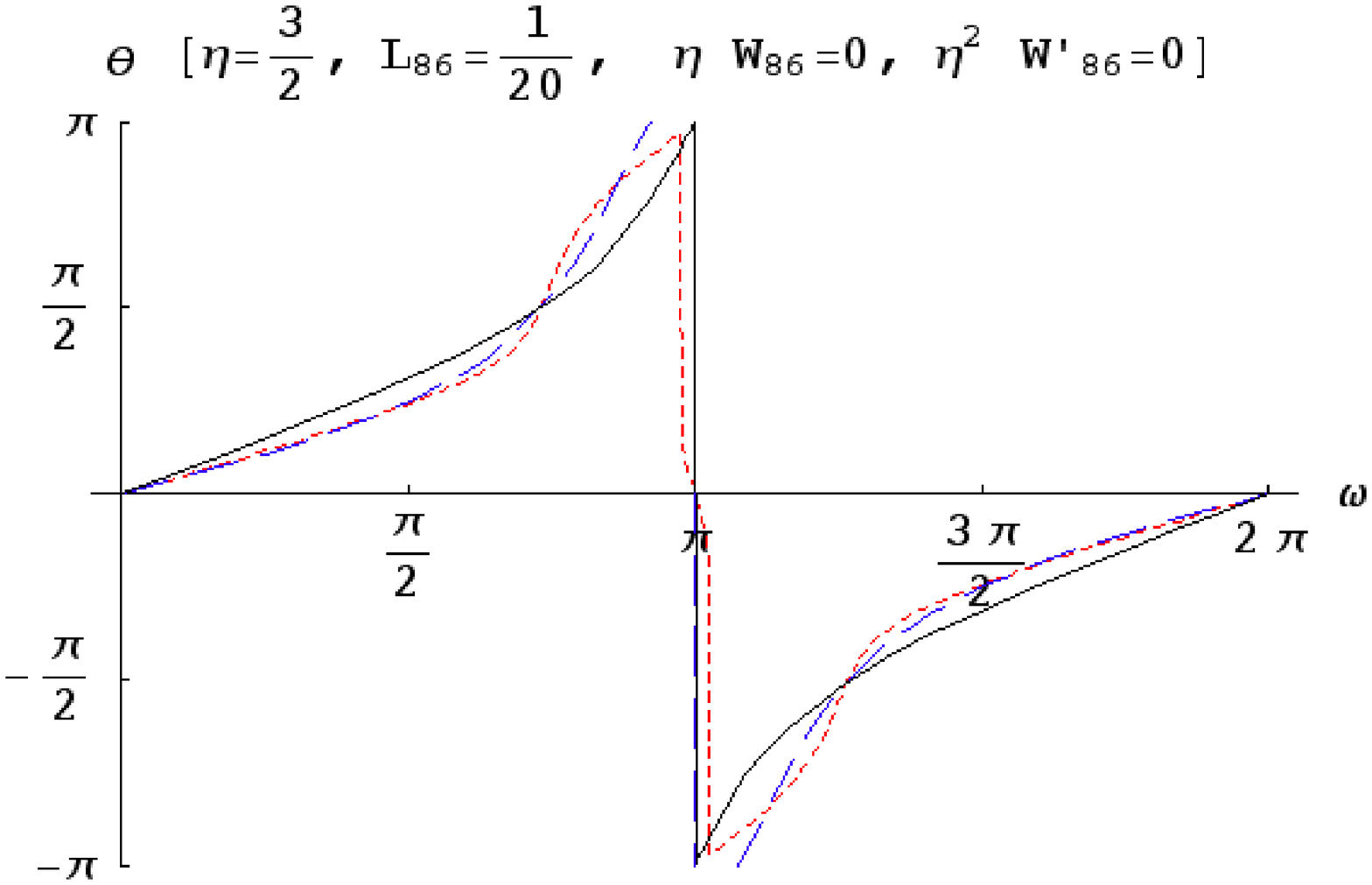}
\includegraphics[height=5cm,width=5cm]{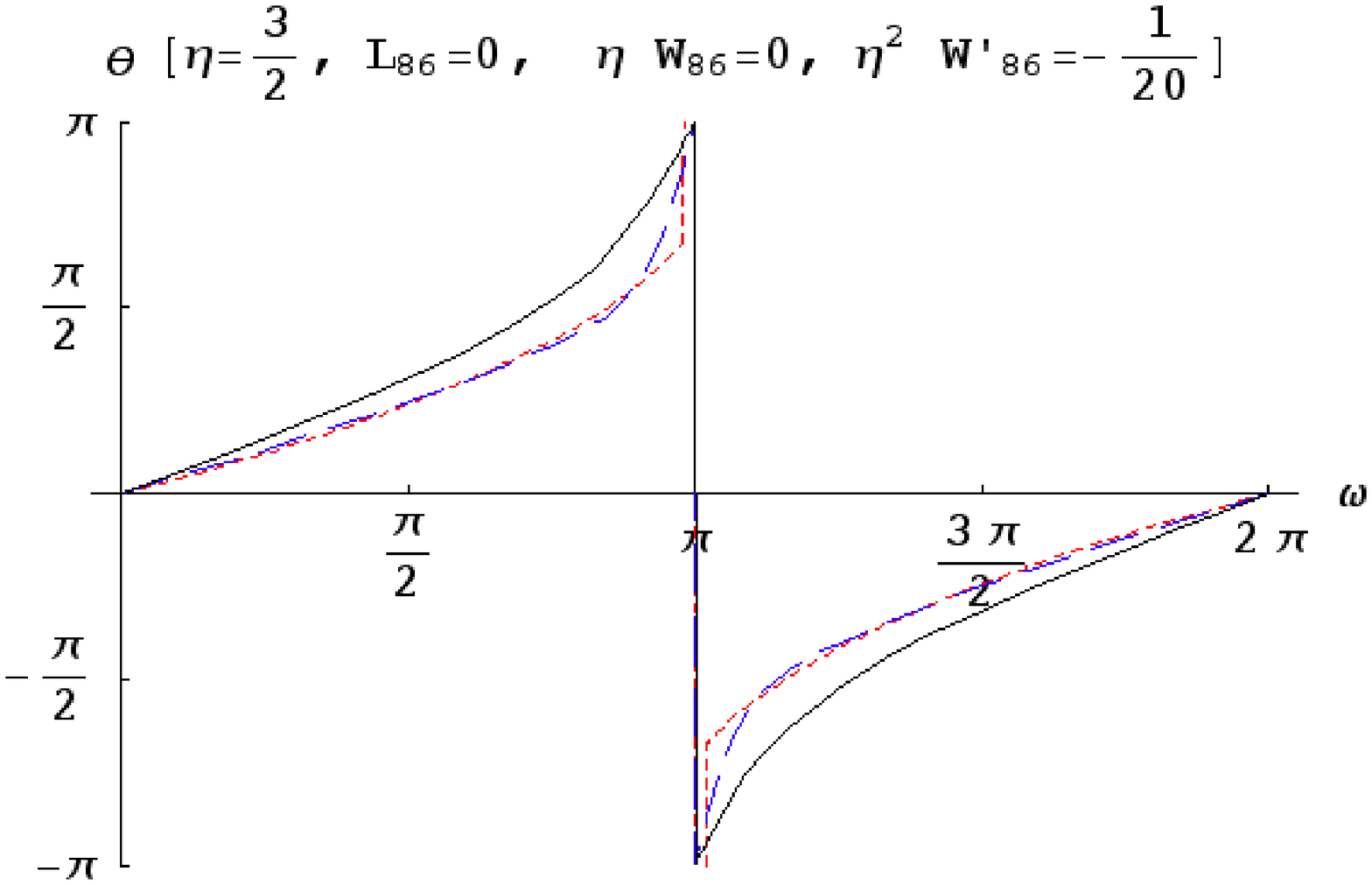}
\includegraphics[height=5cm,width=5cm]{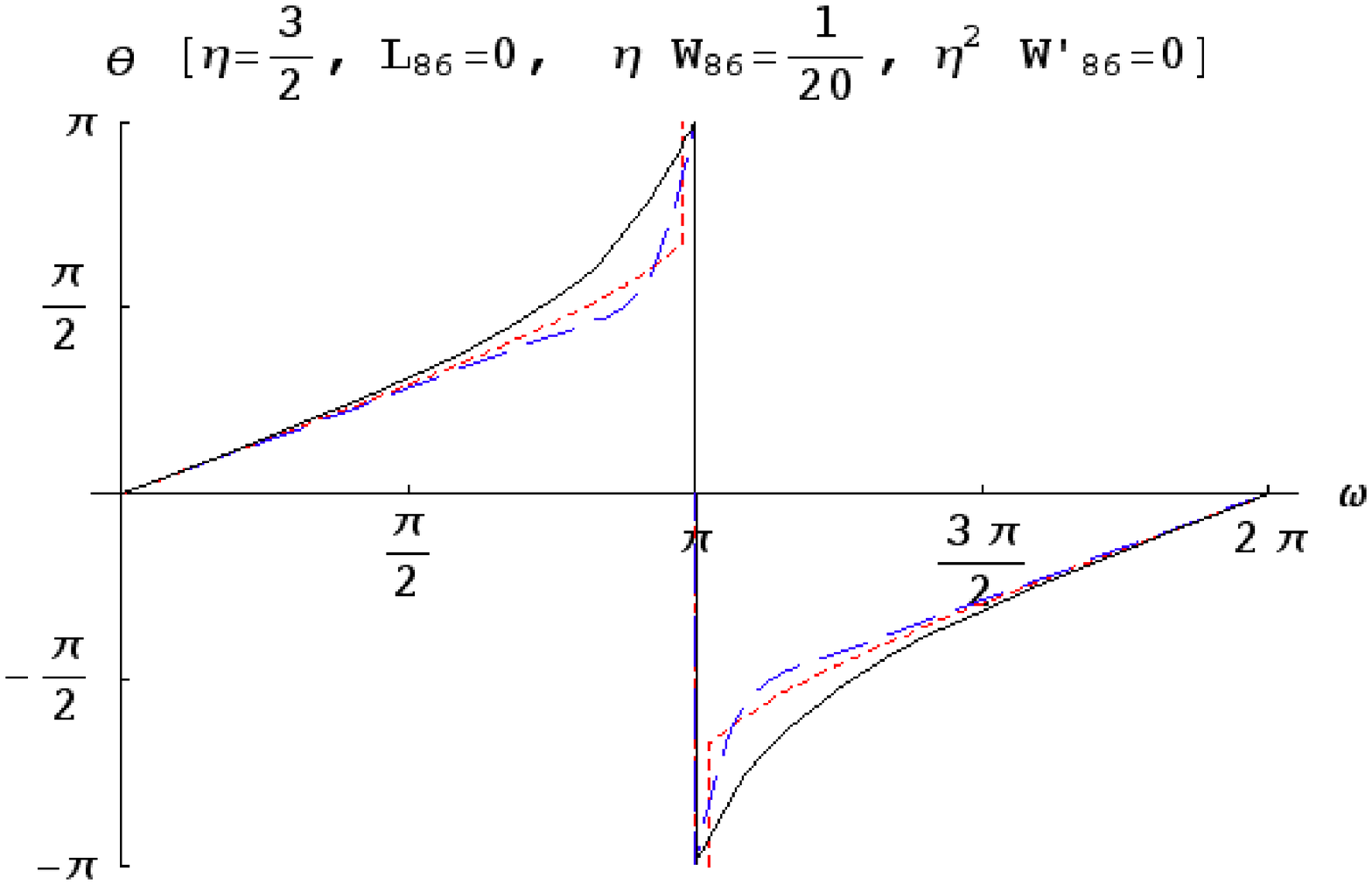}
\includegraphics[height=5cm,width=5cm]{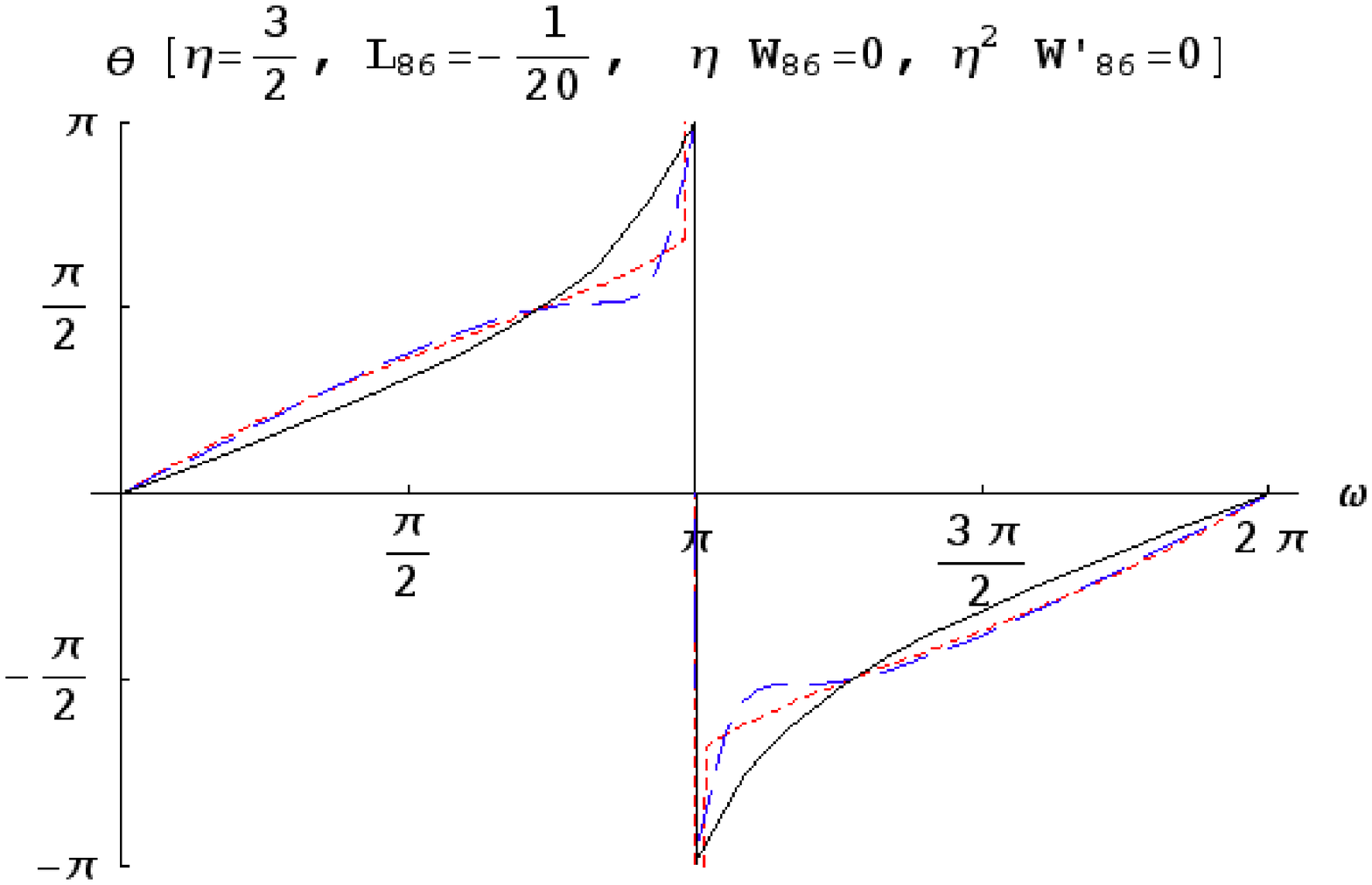}
\caption{\label{fig:NLOvacuum-feta} NLO solution for the vacuum orientation $\theta$ 
as a function of $\omega$ for various choices
of the parameters (dashed-blue). 
NLO solution obtained by numerical minimization of the NLO potential 
(dotted-red).
LO solution (full-black).
}
\end{figure}

\begin{figure}
\includegraphics[height=5cm,width=5cm]{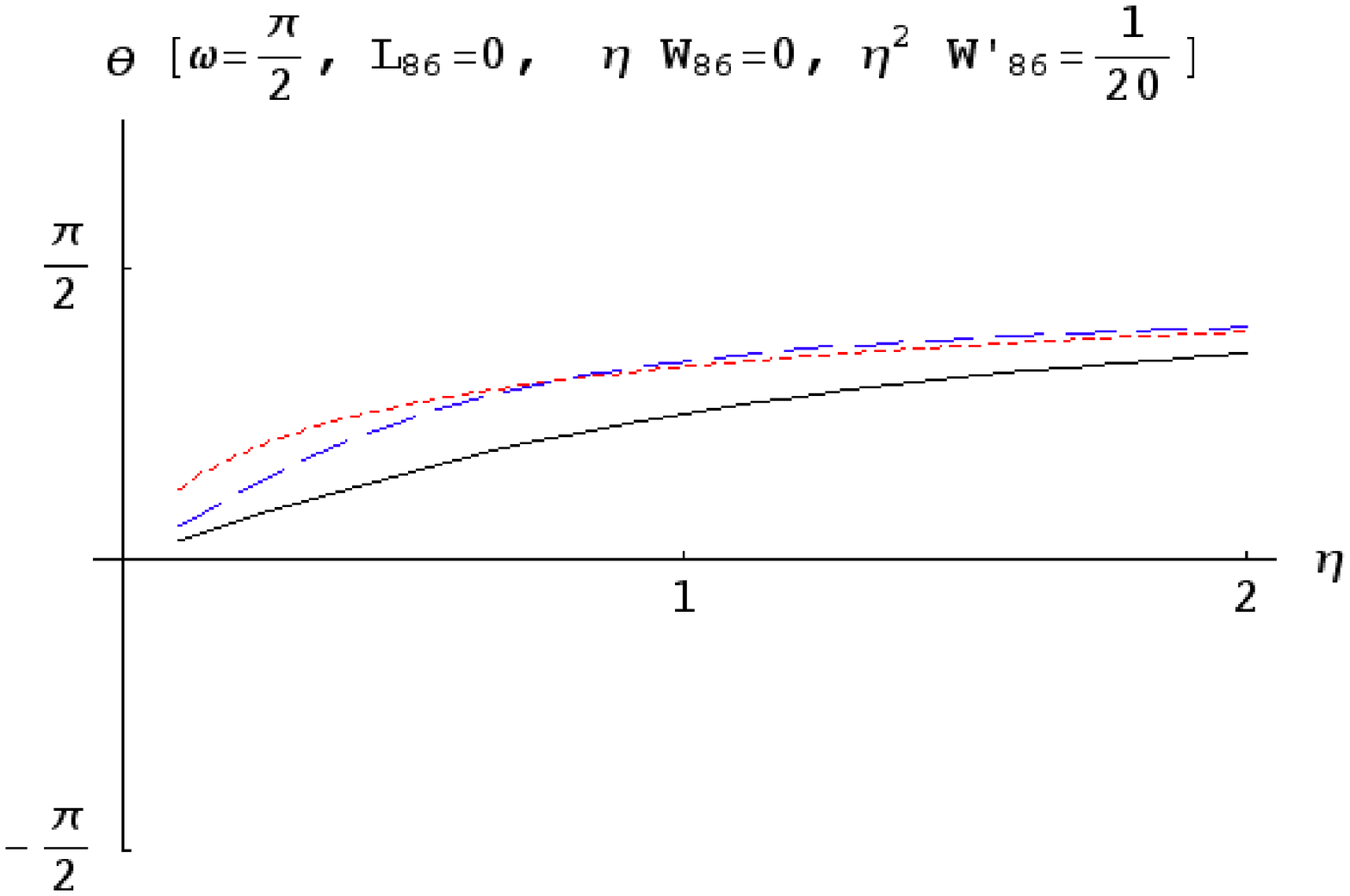}
\includegraphics[height=5cm,width=5cm]{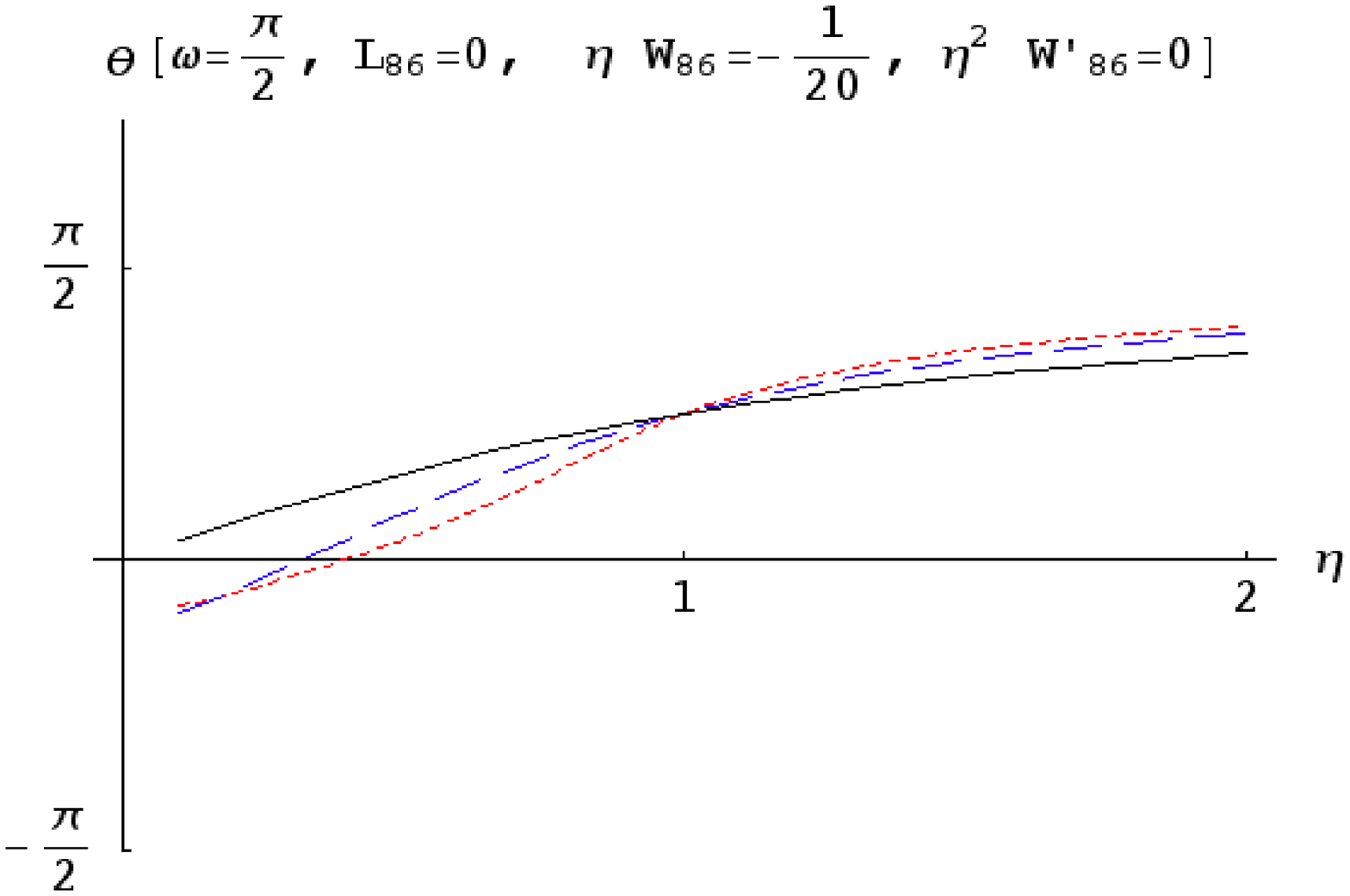}
\includegraphics[height=5cm,width=5cm]{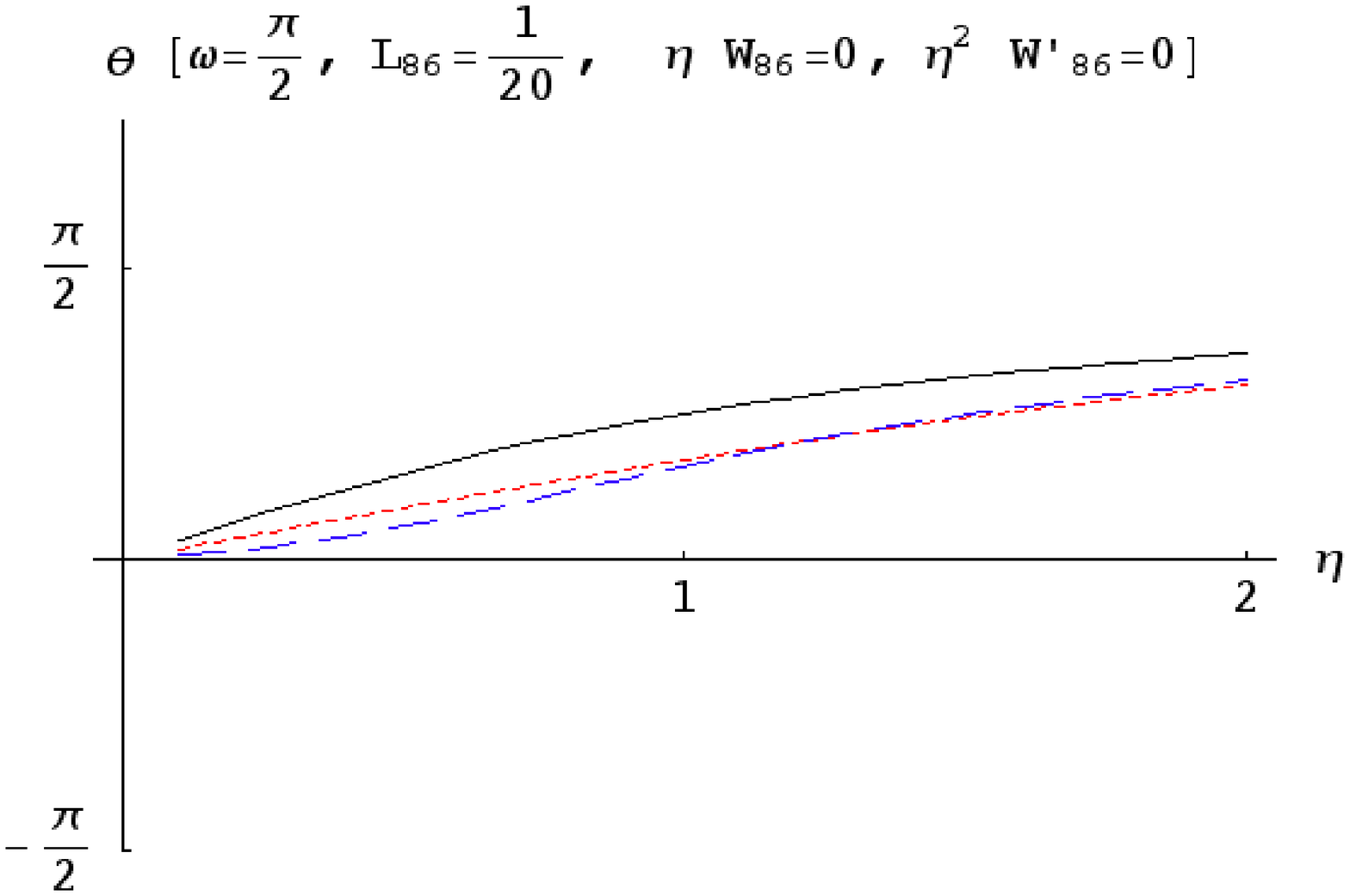}
\includegraphics[height=5cm,width=5cm]{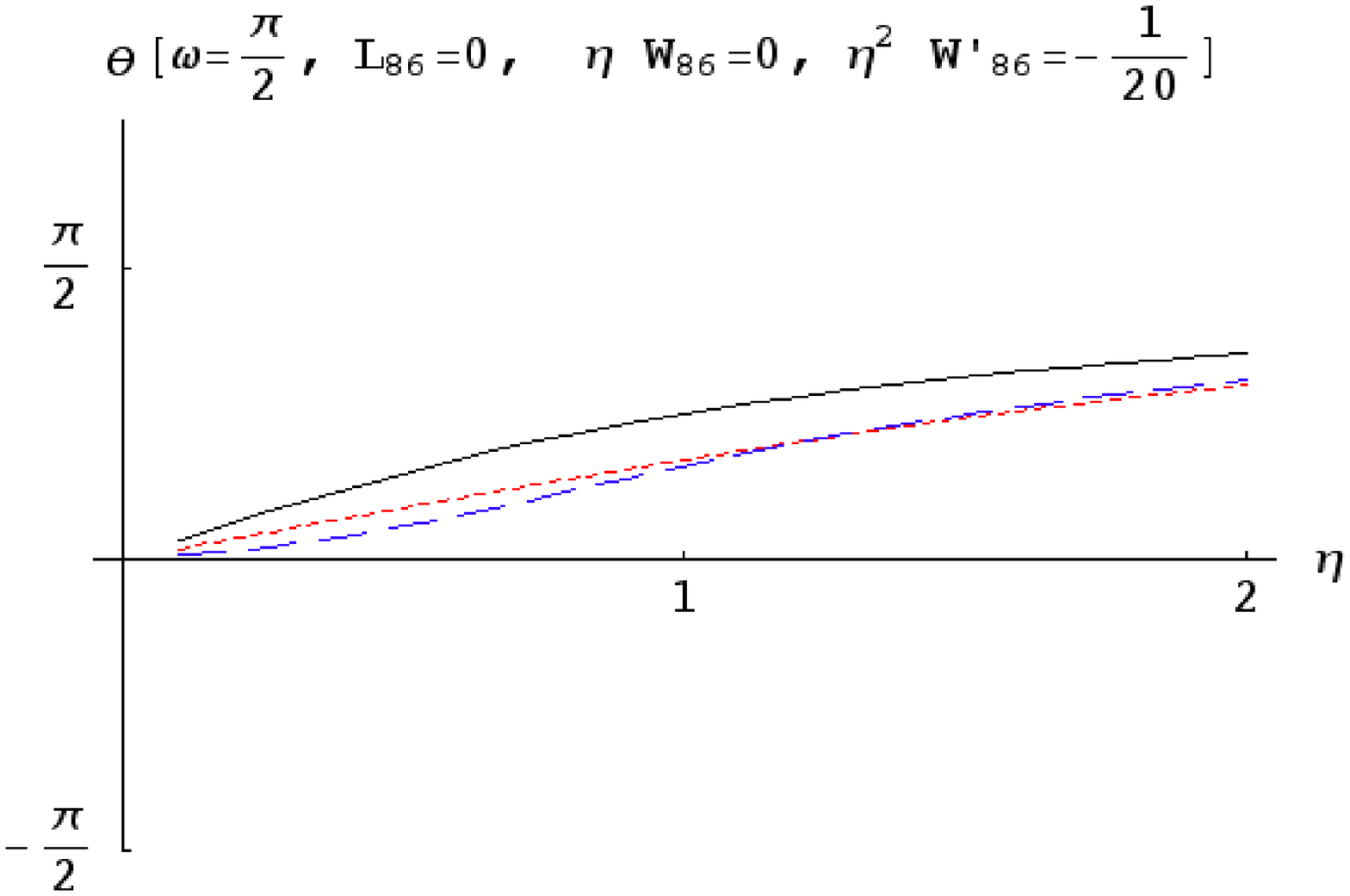}
\includegraphics[height=5cm,width=5cm]{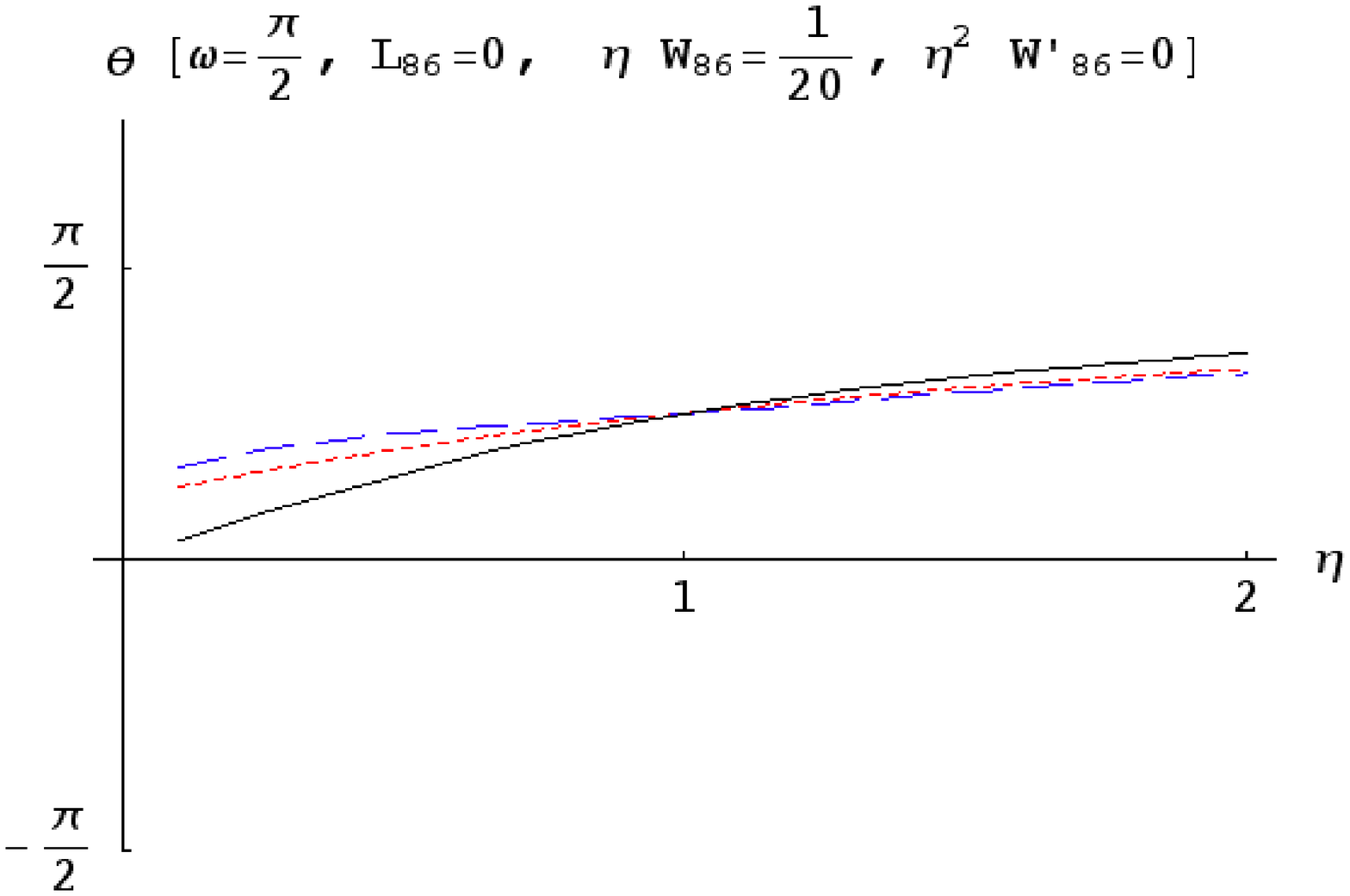}
\includegraphics[height=5cm,width=5cm]{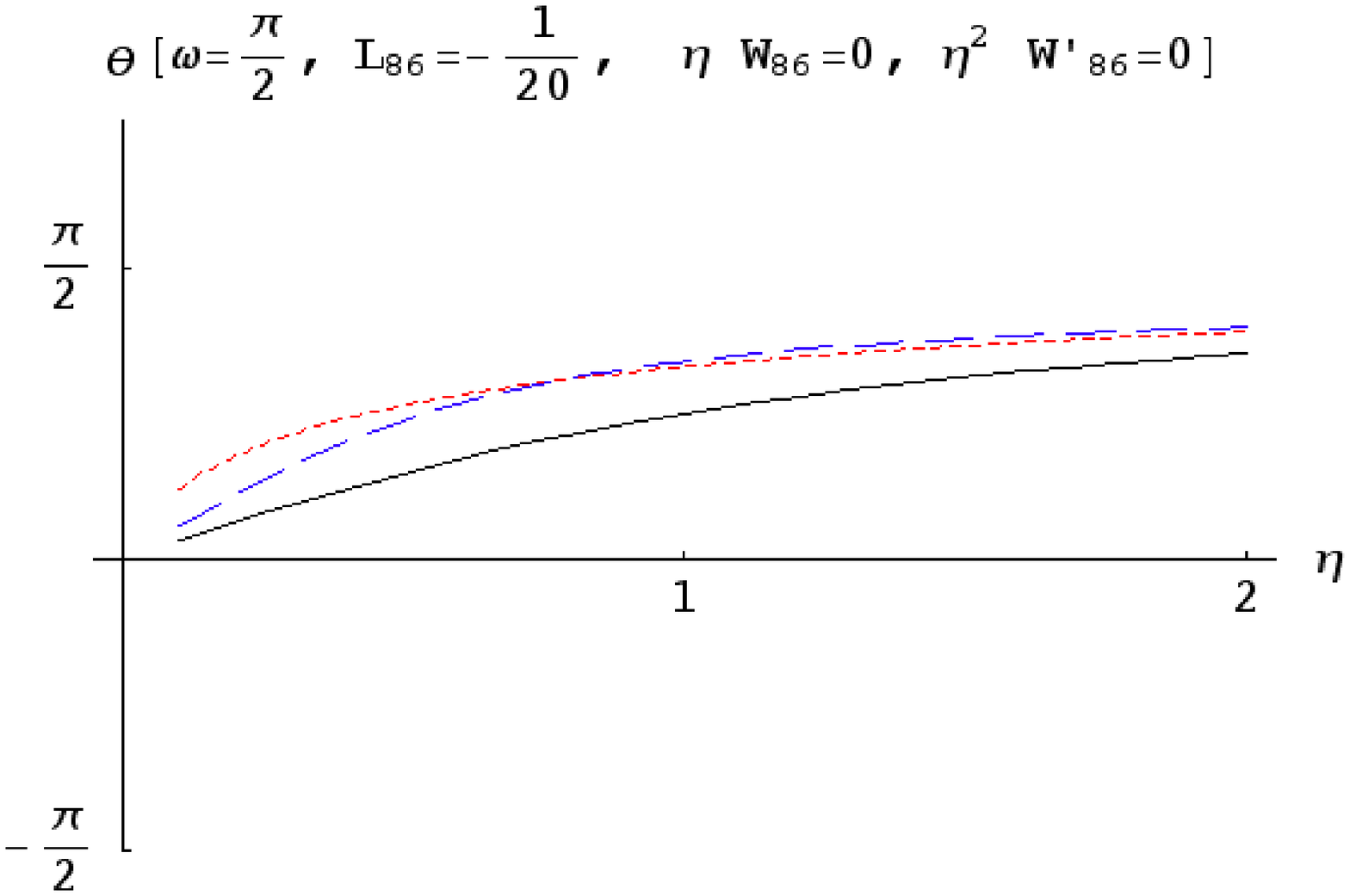}
\includegraphics[height=5cm,width=5cm]{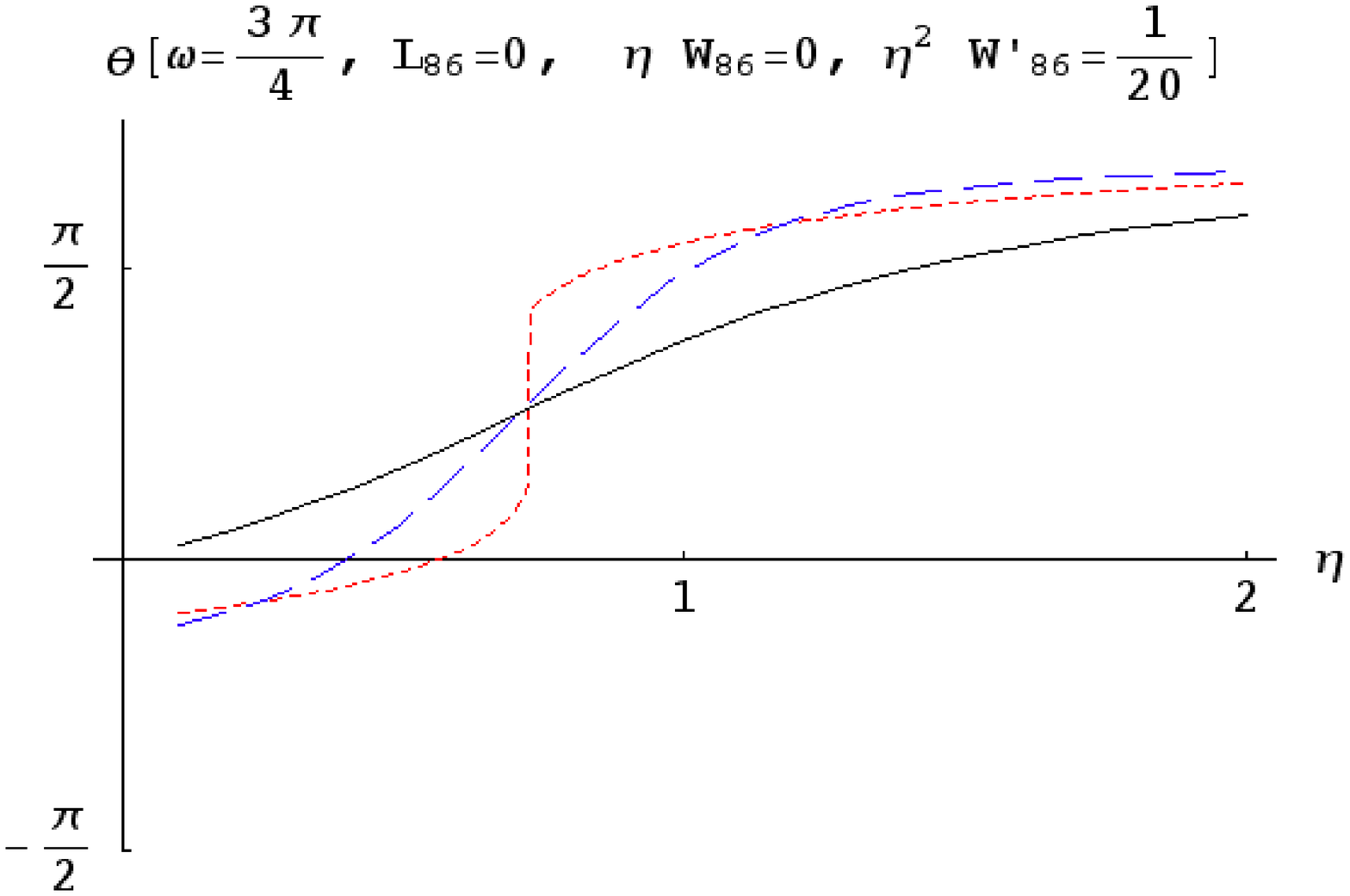}
\includegraphics[height=5cm,width=5cm]{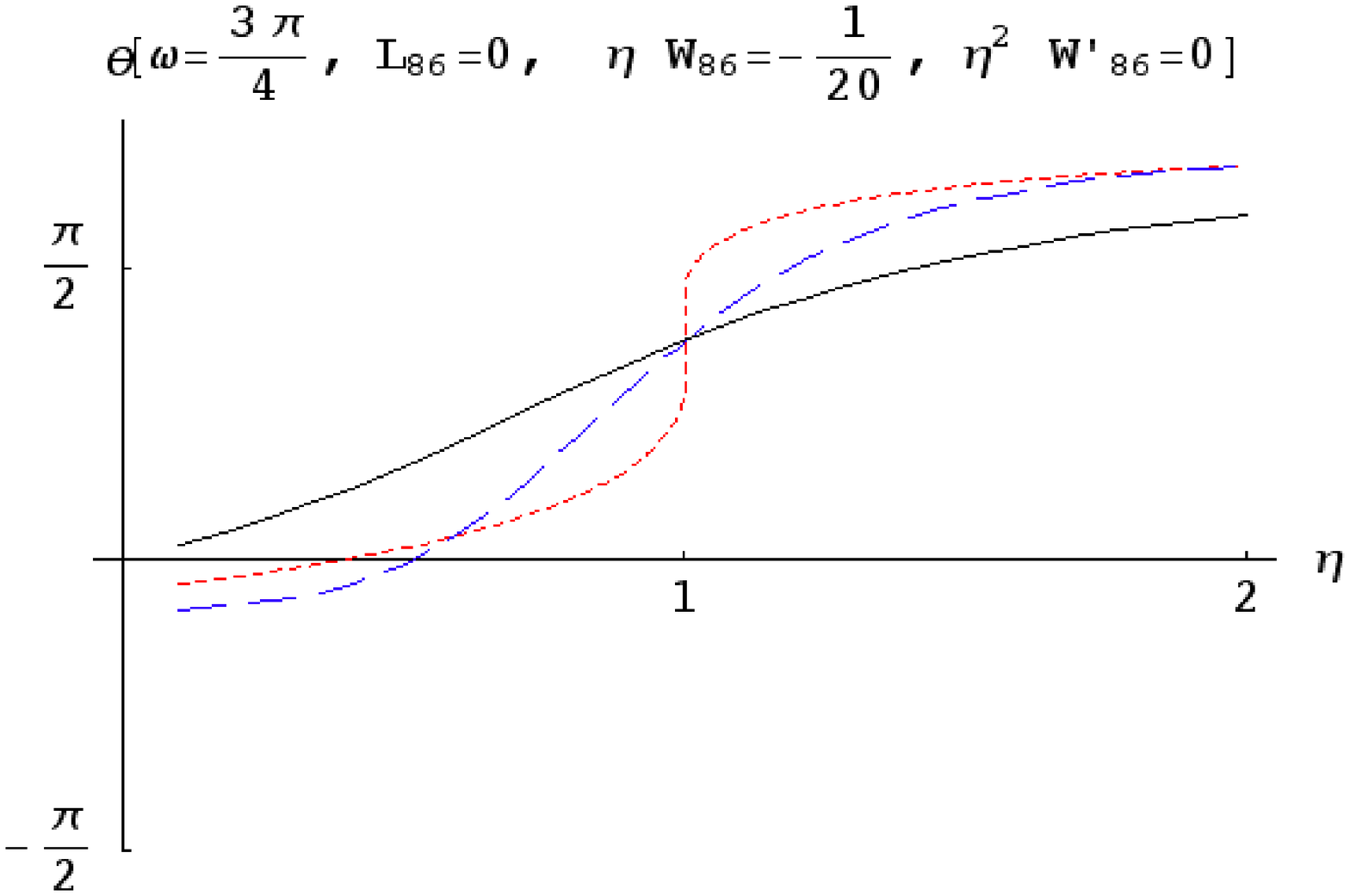}
\includegraphics[height=5cm,width=5cm]{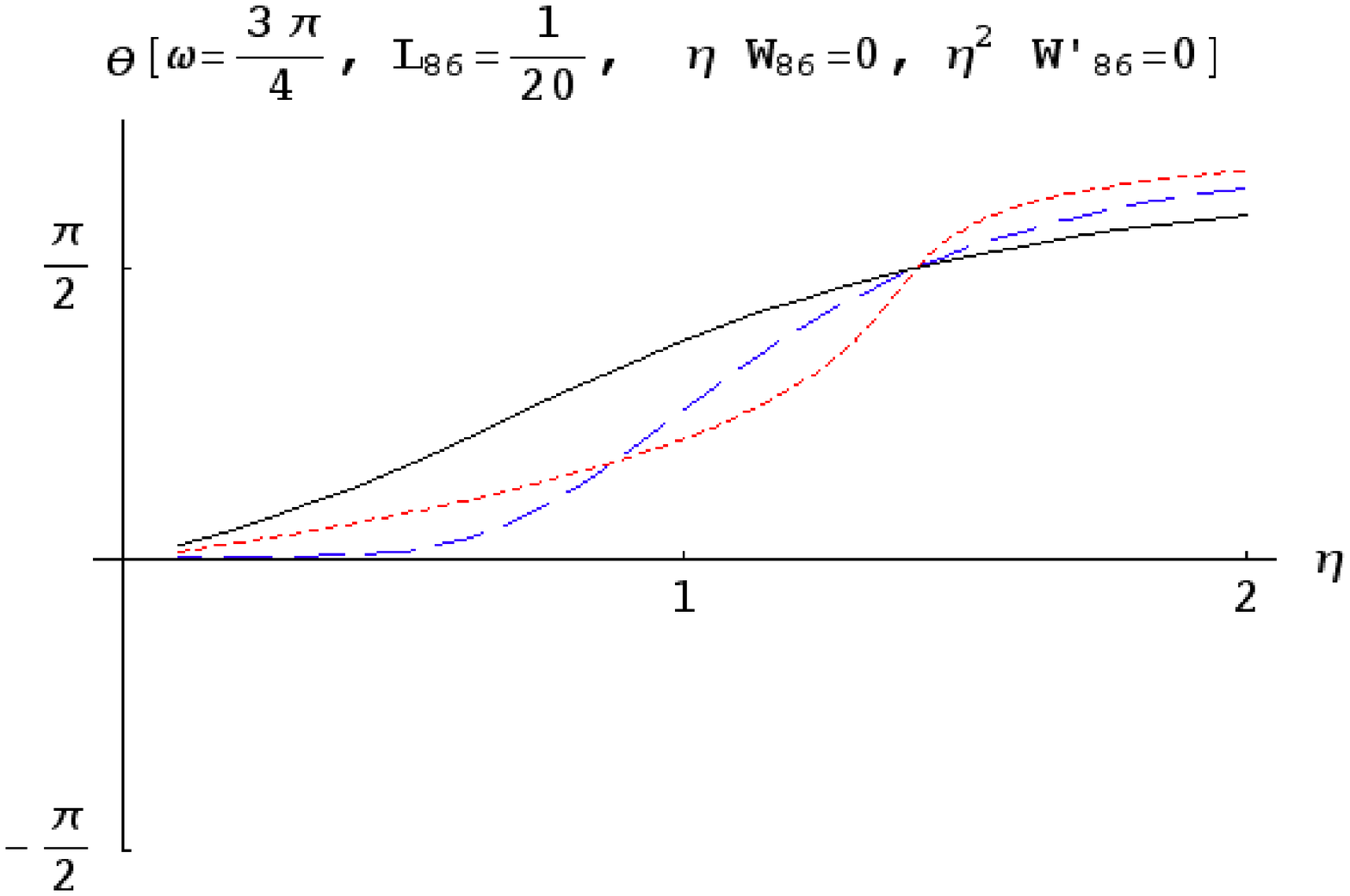}
\includegraphics[height=5cm,width=5cm]{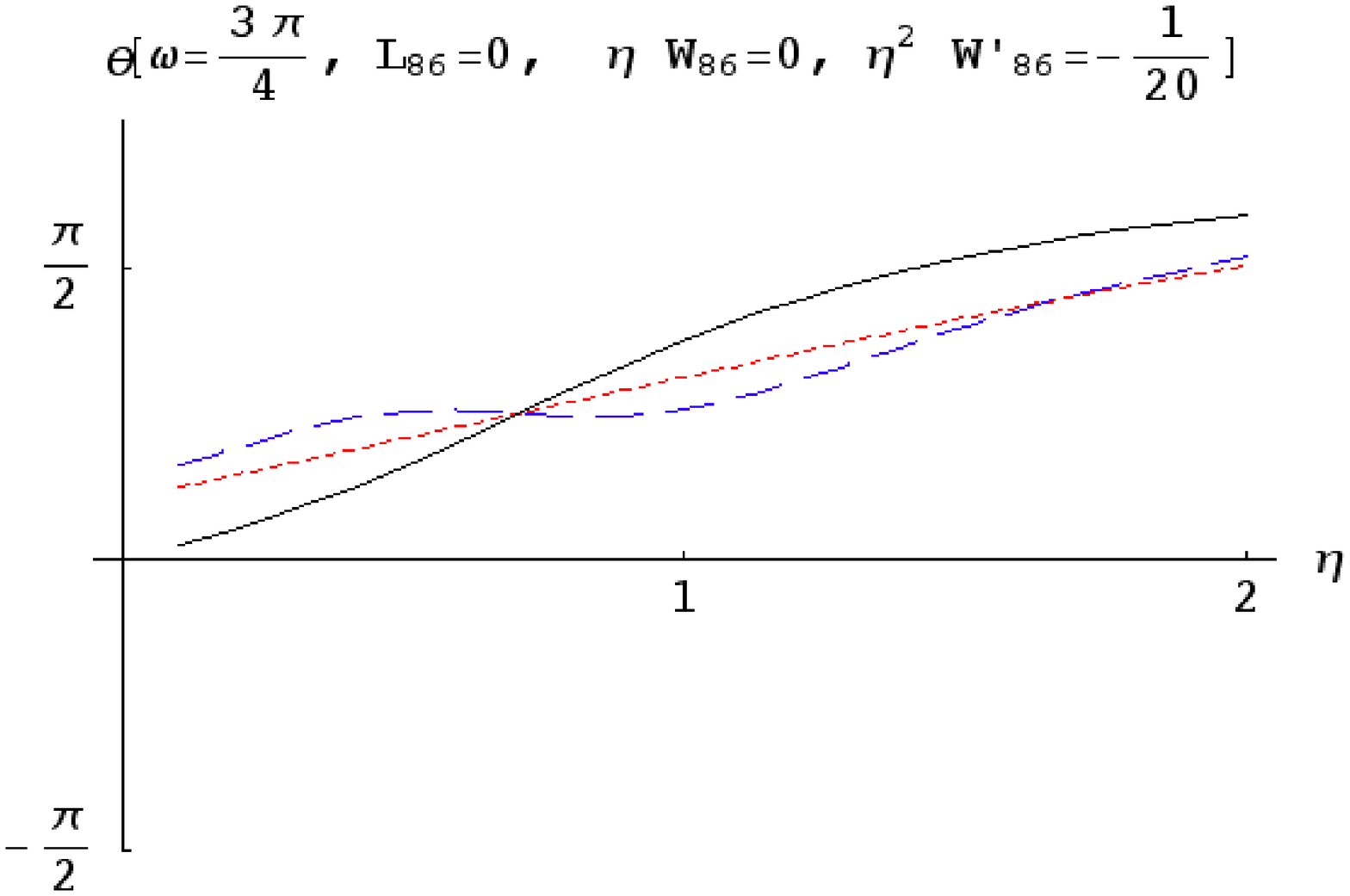}
\includegraphics[height=5cm,width=5cm]{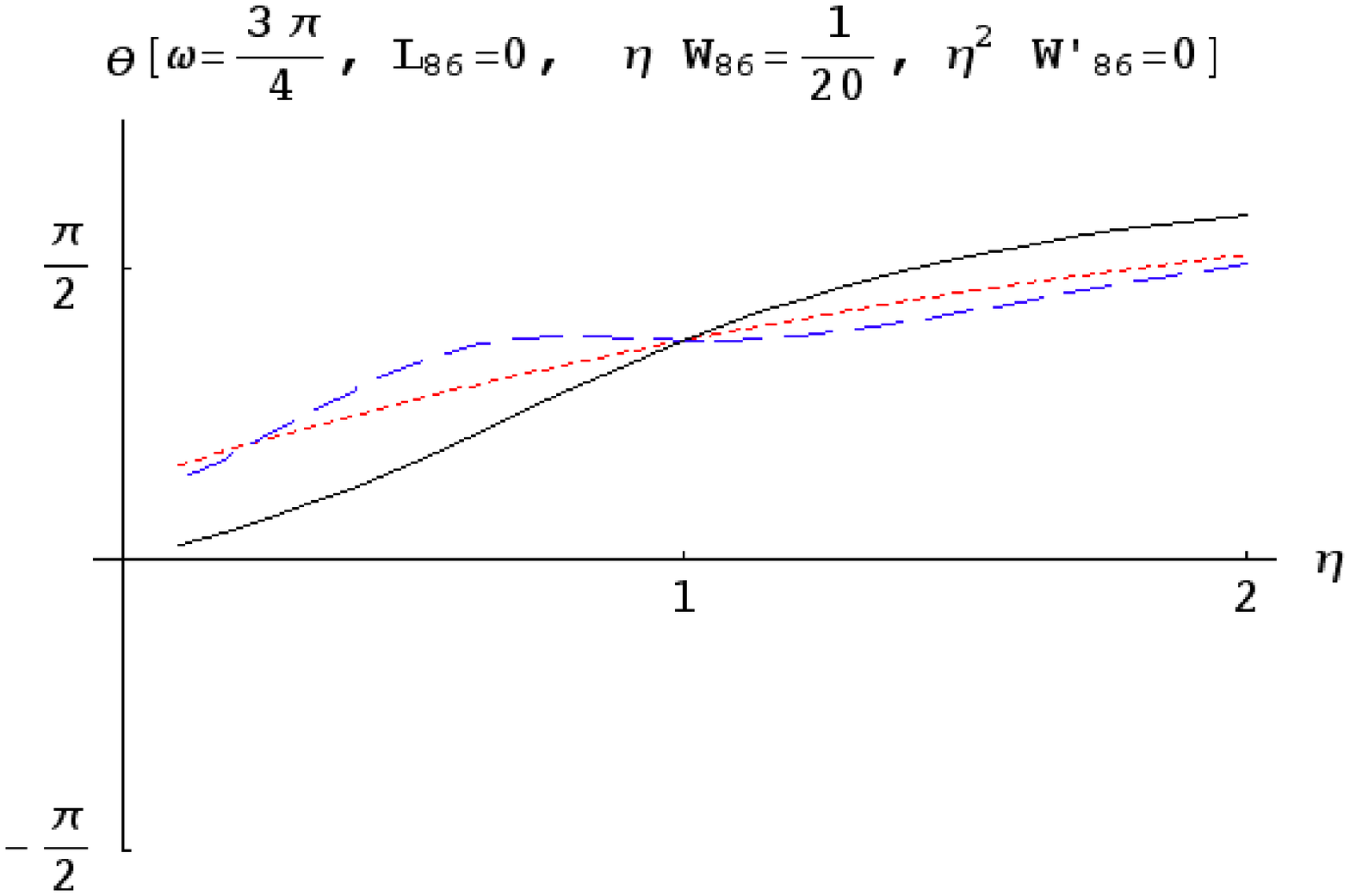}
\includegraphics[height=5cm,width=5cm]{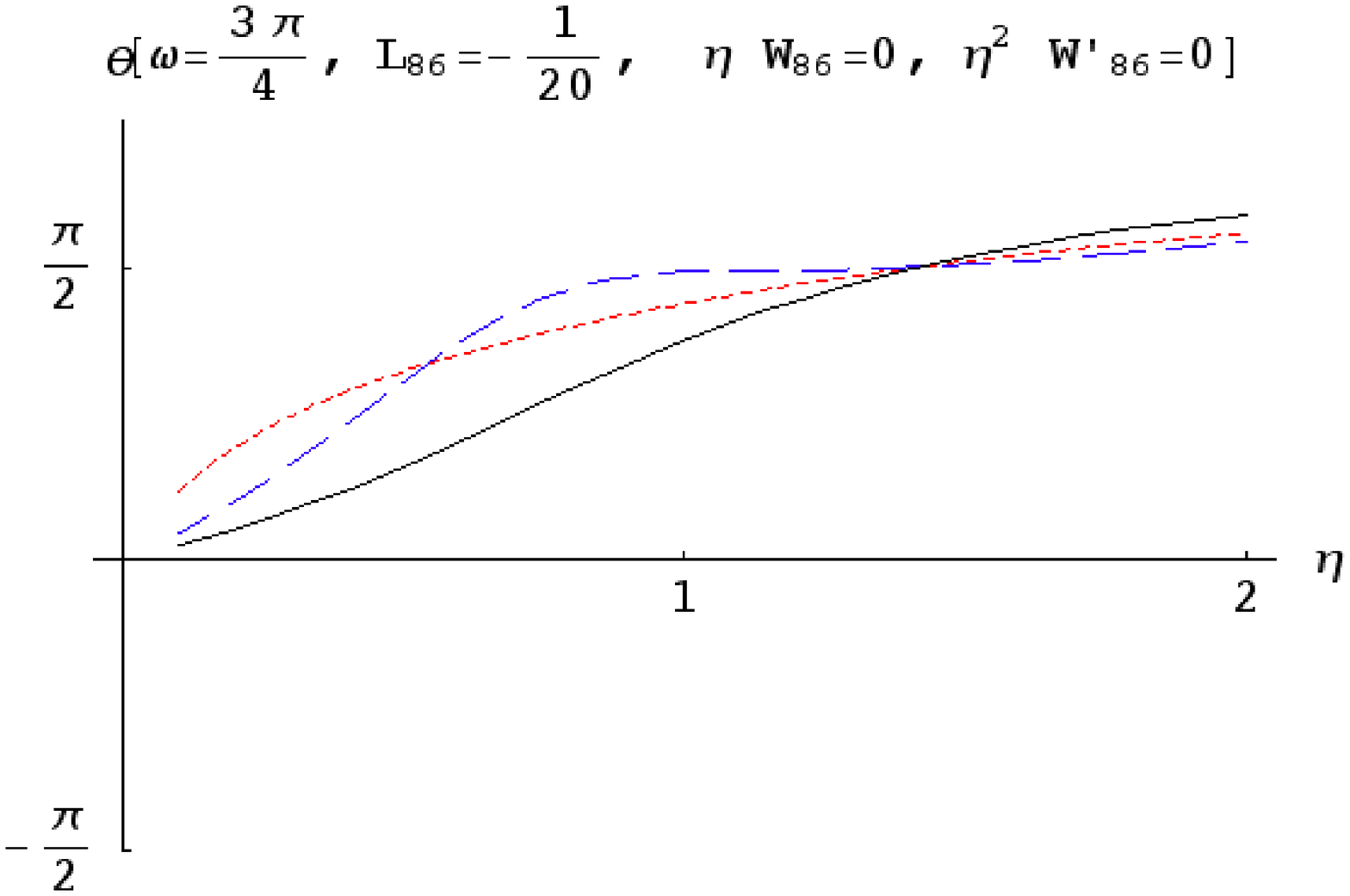}
\caption{\label{fig:NLOvacuum-fth} NLO solution for the vacuum orientation $\theta$ 
as a function of $\eta$ for various choices of the parameters. 
Dashed-blue, dotted-red, full-black as before.
}
\end{figure}

The NLO correction to the solution (\ref{eq:solOmegaLO}) is:
\begin{eqnarray} \label{eq:solOmegaNLO}
\theta&=& \theta_{LO} + 8 \chi ({{\eta }^2}+2 \cos(\omega ) \eta +1)^{-\frac{1}{2}} 
\big[
-
\sin\big(2 {{\tan}^{-1}}\big(\frac{\eta  \sin(\omega )}{\eta  \cos(\omega )+1}\big)\big)
{L_{86}}+ \\
&&+
\sin\big(2 \omega -2 {{\tan}^{-1}}\big(\frac{\eta  \sin(\omega )}{\eta  \cos(\omega )+1}\big)\big)
{{W'}_{86}} {{\eta }^2}
+
\sin\big(\omega -2 {{\tan}^{-1}}\big(\frac{\eta  \sin(\omega )}{\eta  \cos(\omega)+1}\big)\big) 
{W_{86}} \eta 
\big]\nonumber
\end{eqnarray}
In figure \ref{fig:NLOvacuum-feta} we plot (in dashed-blue)  the solutions (\ref{eq:solOmegaNLO}) 
for some typical value of the parameters. The parameter $\eta$ is chosen in order to put
in evidence the effect of lattice artifacts ($\eta=1/3,3/2$), while $\chi$ and the LEC's
are taken in the range that one can expect from typical lattice simulations: 
($\chi L_{86} ,\,  \chi \eta W_{86} ,\, \chi \eta^2 W'_{86} \,=\, \{0, \pm 1/20\}$;
see the discussion at the end of this section).
For comparison, next to the solutions (\ref{eq:solOmegaNLO}),  
we plot in full-black the LO solution (\ref{eq:solOmegaLO}), 
and in dotted-red the solution obtained by numerically minimizing the potential (\ref{eq:potNLO}).
The numerical minimization is performed with the {\em Mathematica} routine {\tt FindMinimum}.
In particular the comparison between the dotted-red and the dashed-blue 
curves provides an interesting 
visualization of what one could expect from NNLO corrections.
Another insight is given by figure \ref{fig:NLOvacuum-fth} in which the same NLO vacuum 
solutions $\theta$ are plotted as a function of $\eta$ for two fixed $\omega=\pi/2, 3\pi/4$. 
We notice that at
$\omega=\pi/2$ the two NLO descriptions are quite consistent over the whole range of $\eta$.
The stability of the solution (\ref{eq:solOmegaNLO}) is still guaranteed by the positivity
of (\ref{eq:doublederiLO}). 
The condition of stability in $\theta$  -- including the NLO contributions is:
\begin{eqnarray}\label{eq:doublederiNLO}
\frac{\partial^2}{\partial \theta^2} {\cal V}_{\chi}&=
{\sqrt{{{\eta }^2}+2  \cos(\omega )  \eta +1}}+ 
\frac{16 \, \chi}{{{\eta }^2}+2  \cos(\omega )  \eta +1}
[(\cos(2  \omega )  {{\eta }^2}+2  \cos(\omega)  \eta +1)  {L_{86}} \\
&+\eta   ((\cos(\omega )  {{\eta }^2}+2  \eta +\cos(\omega ))  {W_{86}}
+\eta   ({{\eta}^2}+2  \cos(\omega )  \eta +\cos(2  \omega ))  {W'_{86}})].\nonumber
\end{eqnarray}
However the LO term go to zero when 
$\eta \rightarrow 1$ and $\omega \rightarrow \pi$ at the same time. 
This is the point
where something new can happen, depending on NLO terms. This is the subject of the next paragraph. 

\subsection{Critical region}
\label{sec:CritReg}
As already justified we can limit ourselves to consider the potential
at $\phi=0$:
\begin{equation} \label{eq:potnoTh}
{\cal V}_{\chi}=
-\cos(\theta )
-\cos(\omega -\theta )  \eta 
-4  \chi [  
\cos(2  \theta )  {L_{86}}
+\cos(\omega -2  \theta )  {W_{86}}  \eta 
+\cos(2  (\omega -\theta ))  {W'_{86}}  {{\eta }^2}
]
\end{equation}
Notice that NLO terms have a periodicity in $\theta$ which is double as fast as the one
in the LO terms (i.e. they have period $\pi$ instead of $2 \pi$).  
This means that the NLO terms can introduce at most a new pair of stationary
points (in this case a new minimum and a new maximum). In more formal terms, the 
condition $\frac{\partial}{\partial \theta} {\cal V}_{\chi LO}=0$ is a polynomial 
of degree 2 in the complex variable $z=e^{i \theta}$, while the same equation for
(\ref{eq:potnoTh}) has degree 4. The appearance of new stationary points is 
actually only possible if the NLO terms are strong enough to compensate the LO term
in the stability condition (\ref{eq:doublederiNLO}). Since we always assume that $\chi X$ 
($X=$ any LEC) are
small, the only possibility is when $(\eta^2 + 2 \eta \cos(\omega) + 1)$ is small, 
i.e. when both
$\eta \rightarrow 1$ and $\omega \rightarrow \pi$. 

In order to describe this region we choose a convenient  parameterization:
\begin{equation} \label{eq:paramcrit}
\eta = 1 + \chi \delta\eta, 
\;\;\;\;\;\;\;\;\;\; 
\omega = \pi + \chi \delta\omega, 
\end{equation}
and we assume that $\delta\omega$ and $\delta\eta$ are dimensionless number of the same order 
of magnitude as the LEC's. Roughly speaking
one can think of $\delta \eta$ as a very small
ordinary mass, and $\delta\omega$ as a very small twisted mass.
In these variables the potential (at first order in $\chi$) becomes:
\[
{\cal V}_{\chi}/{\chi} = 
\delta \eta \cos (\theta)
-8 (L_{86} - W_{86} + W'_{86}) \cos (\theta)^2
+ \delta \omega \sin (\theta)
+ O(\chi)
\]
If $\delta\omega=0$, we recover the familiar potential first studied in \cite{Sharpe:1998xm}.
This identifies the coefficients $c_1$, $c_2$ and $\epsilon$ 
introduced in \cite{Sharpe:1998xm} with
\begin{eqnarray} \label{eq:ident}
c_2 &=& -8 (L_{86} - W_{86} + {W'}_{86}), \\
-c_1&=&\delta\eta = F^2 \frac{W_0 a - B_0 m}{2 (B_0 m)^2}, \nonumber\\
\epsilon &=&c_1/(2 c_2) =  
F^2 \frac{W_0 a - B_0 m}{32 (B_0 m)^2 [(L_{86} - W_{86} + {W'}_{86})]} \nonumber
\end{eqnarray}
Of course an overall (positive) factor in $c_1$ and $c_2$ is irrelevant. 

It is interesting to see that, although some LEC's depend on the definition of the mass,
the picture above is stable under such redefinition. We mentioned in section \ref{sec:Lag}
that one can redefine the mass as in (\ref{eq:matrans}) and go to very small $m^* \sim a^2$.
In this setup one expects that only the $O(a^2)$ LEC's $W'$ contribute (see \cite{Sharpe:1998xm}). 
However from (\ref{eq:LECtrans}) one finds $(L_{86} - W_{86} + {W'}_{86})= {W'}^*_{86}$,
showing  perfect consistency  between the two descriptions. We can also observe
that the combination $(L_{86} - W_{86} + {W'}_{86})$ is invariant under renormalization
(see section \ref{sec:Masses}). The constant ${W'}^*_{86}$, instead, does not
renormalize because there is no LO term in $a^*$ which produces divergences which need
to be subtracted.

Before introducing a non zero $\delta\omega$, 
we recall the results of the analysis of \cite{Sharpe:1998xm}:
\begin{enumerate}
\item if $c_1>0$ (positive mass $m^*$) 
and $2 c_2 < c_1$ (i.e. $[L_{86} - W_{86} + {W'}_{86}]$ positive or small negative), 
the solution is $\theta=0$. 
\item if $c_1>0$ (positive mass $m^*$) 
and $c_1 < 2 c_2$ (i.e. $[L_{86} - W_{86} + {W'}_{86}]$ large  negative), 
the solution is $\cos(\theta)=\epsilon>0$
(the chiral condensate has a positive component like in 1.). 
\item if $c_1<0$ (negative mass $m^*$) 
and $2 c_2 < -c_1$ (i.e. $[L_{86} - W_{86} + {W'}_{86}]$ positive or small negative), 
the solution is $\theta=\pi$. 
\item if $c_1<0$ (negative mass $m^*$) 
and $-c_1 < 2 c_2$ (i.e. $[L_{86} - W_{86} + {W'}_{86}]$ large  negative), 
the solution is $\cos(\theta)=\epsilon<0$
(the chiral condensate has a negative component like in 3.). 
\end{enumerate}

\begin{figure}
\includegraphics[height=5cm,width=5cm]{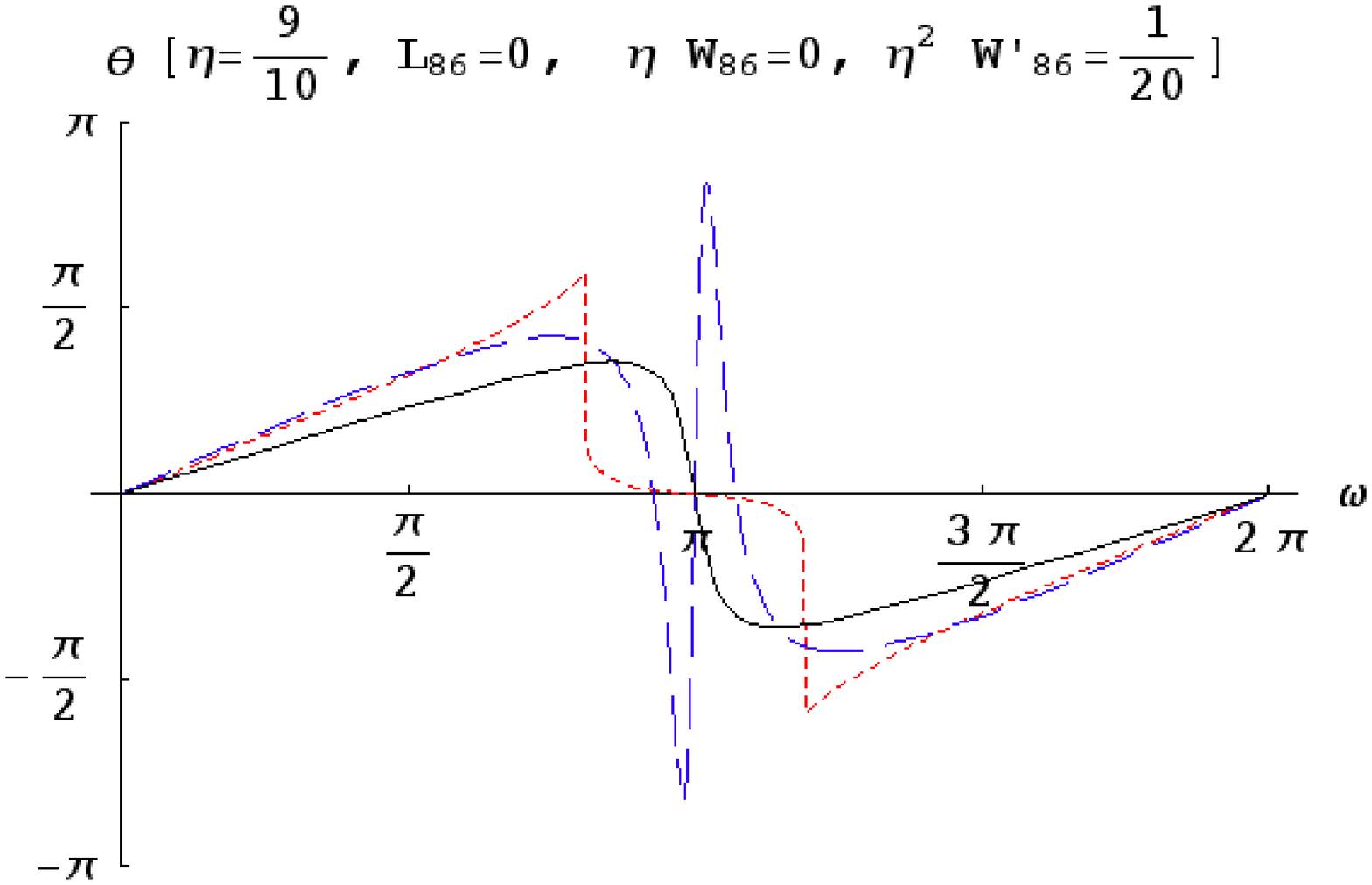}
\includegraphics[height=5cm,width=5cm]{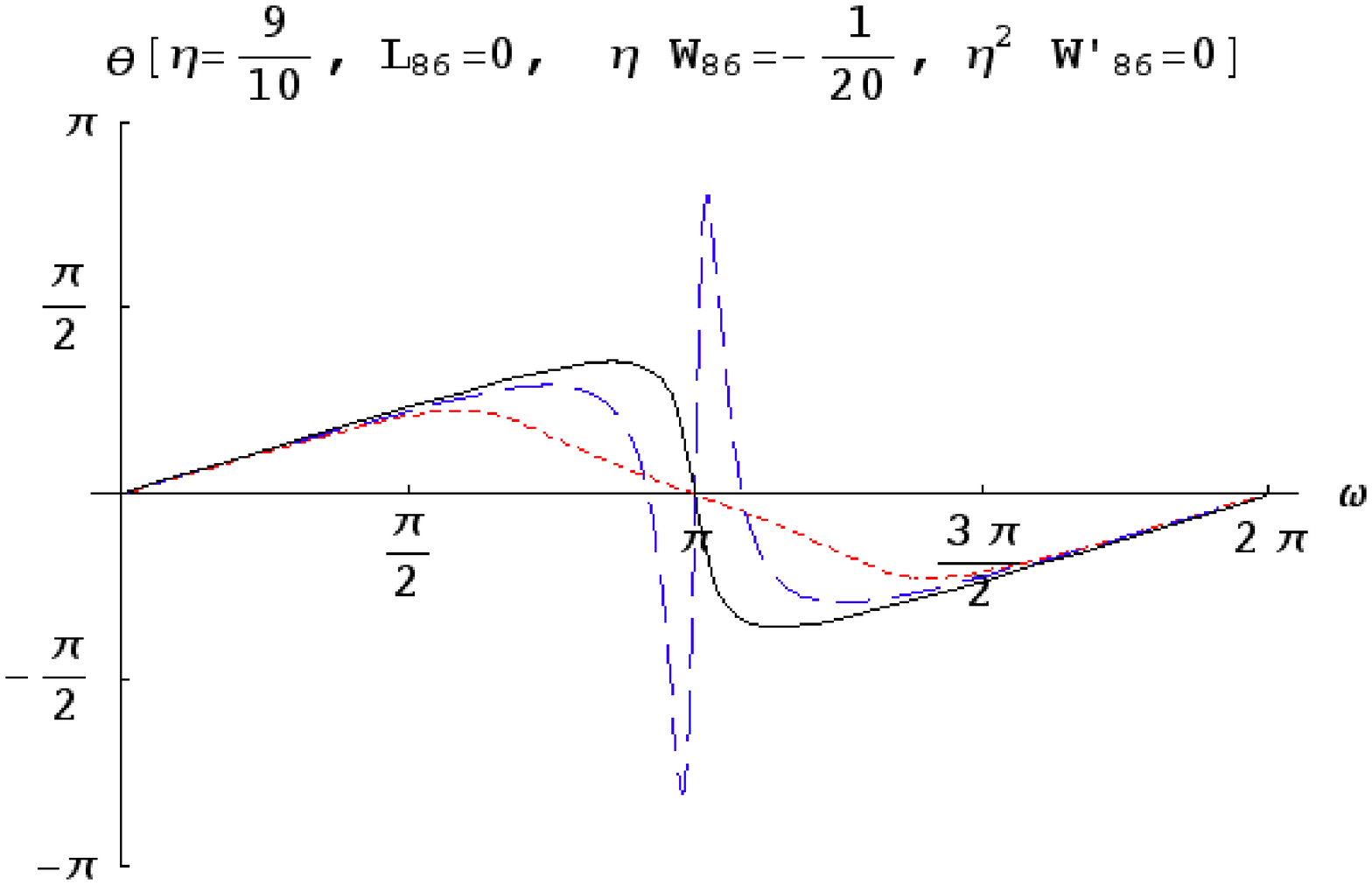}
\includegraphics[height=5cm,width=5cm]{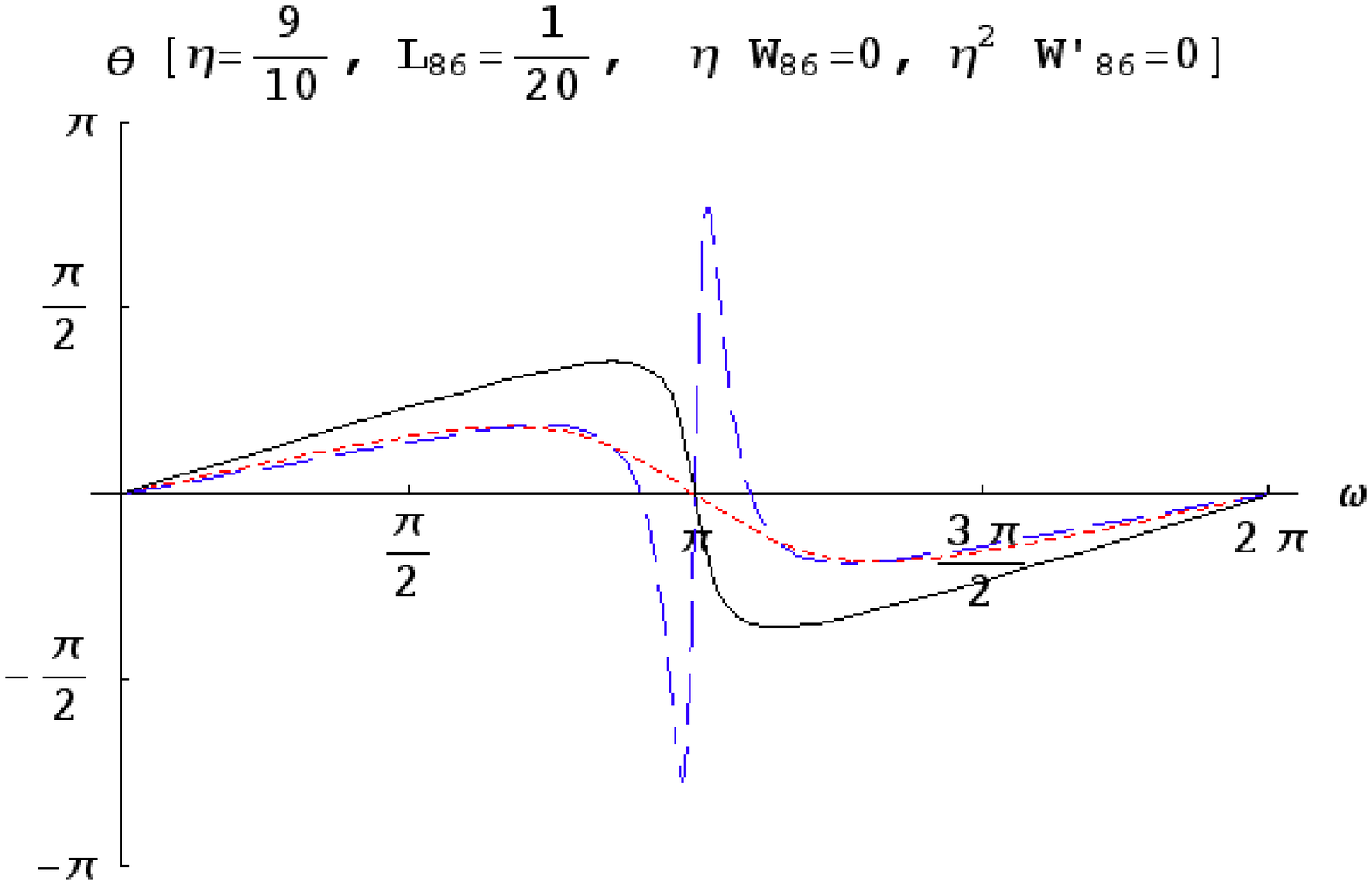}
\includegraphics[height=5cm,width=5cm]{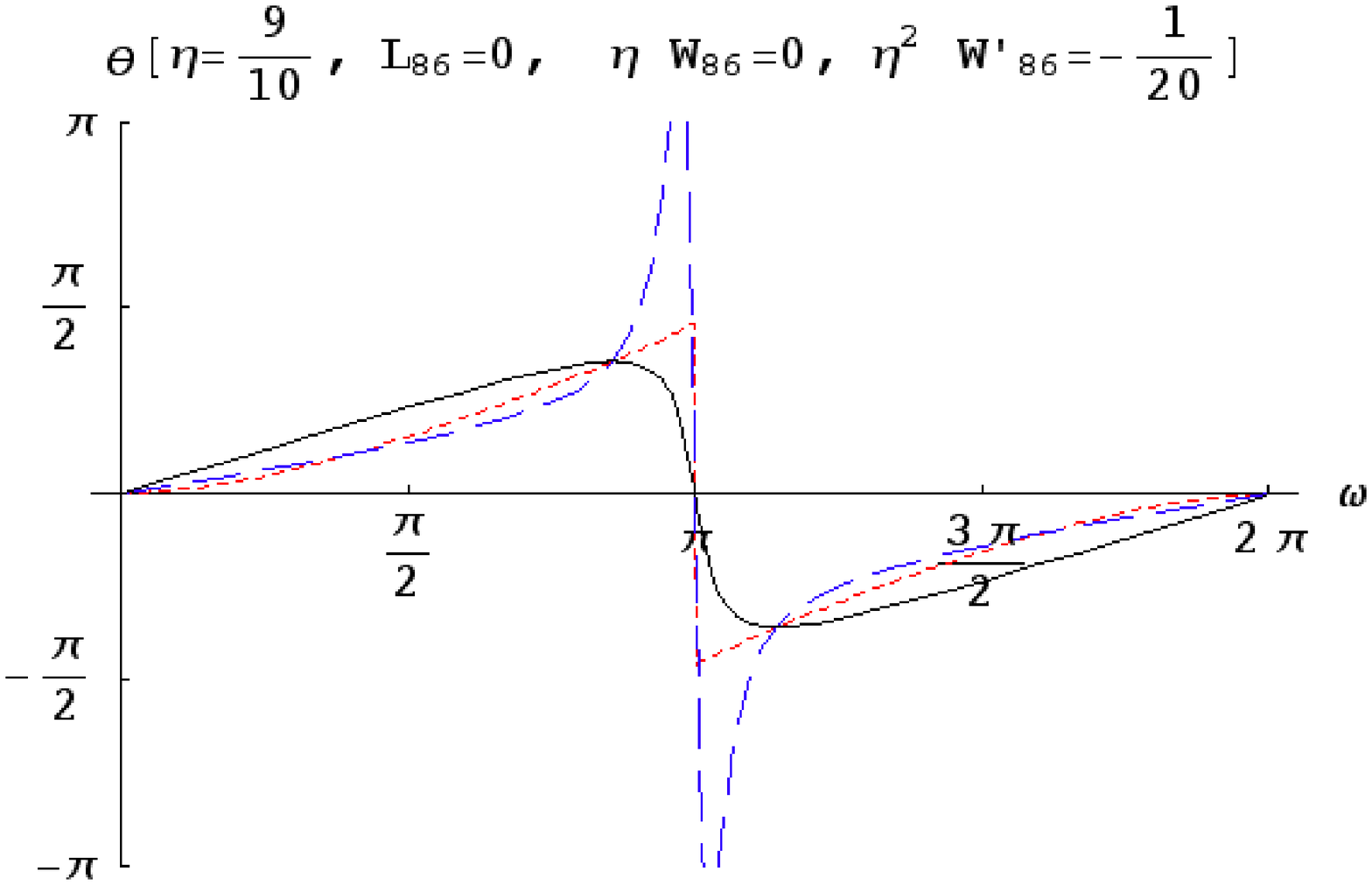}
\includegraphics[height=5cm,width=5cm]{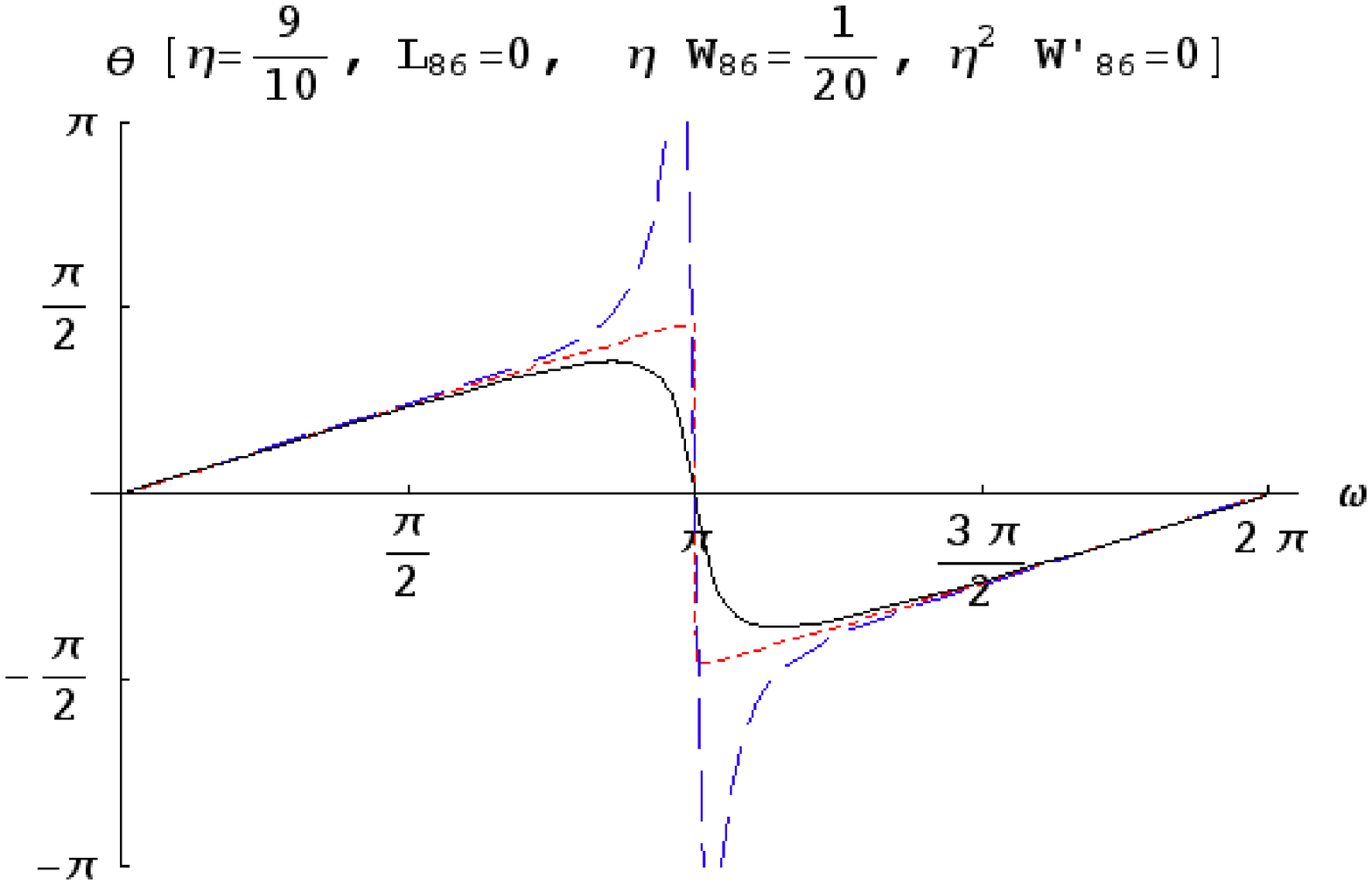}
\includegraphics[height=5cm,width=5cm]{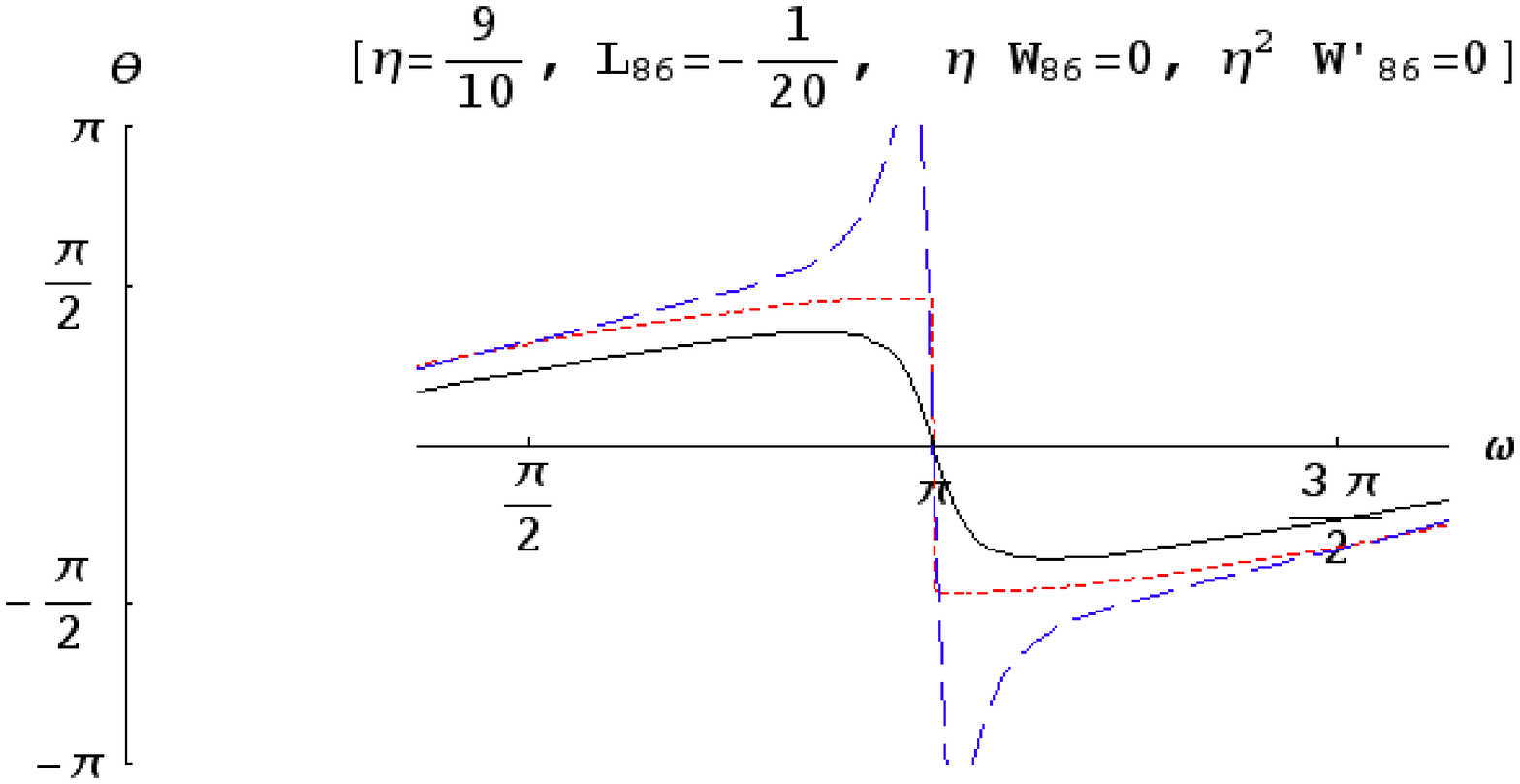}
\includegraphics[height=5cm,width=5cm]{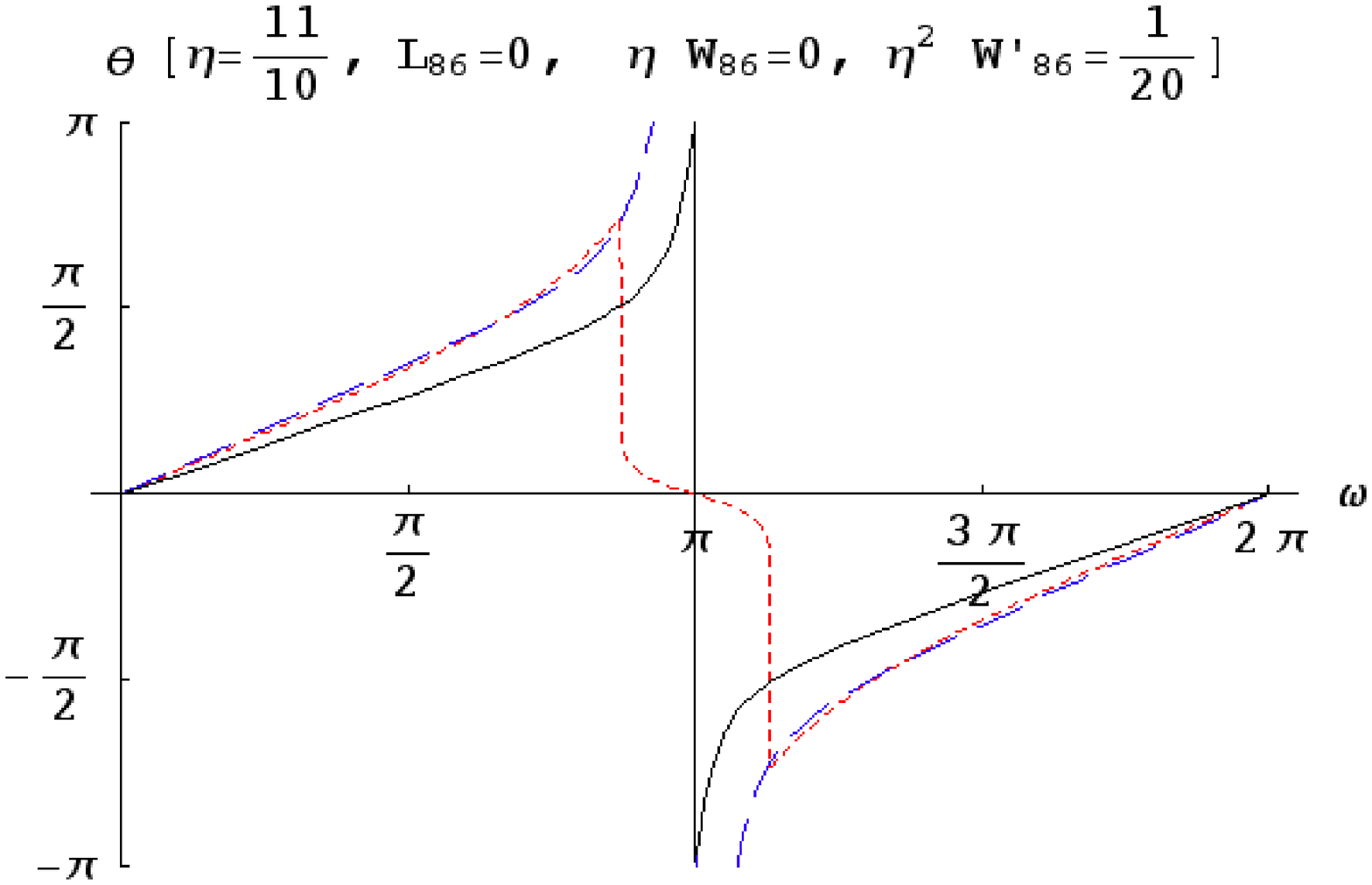}
\includegraphics[height=5cm,width=5cm]{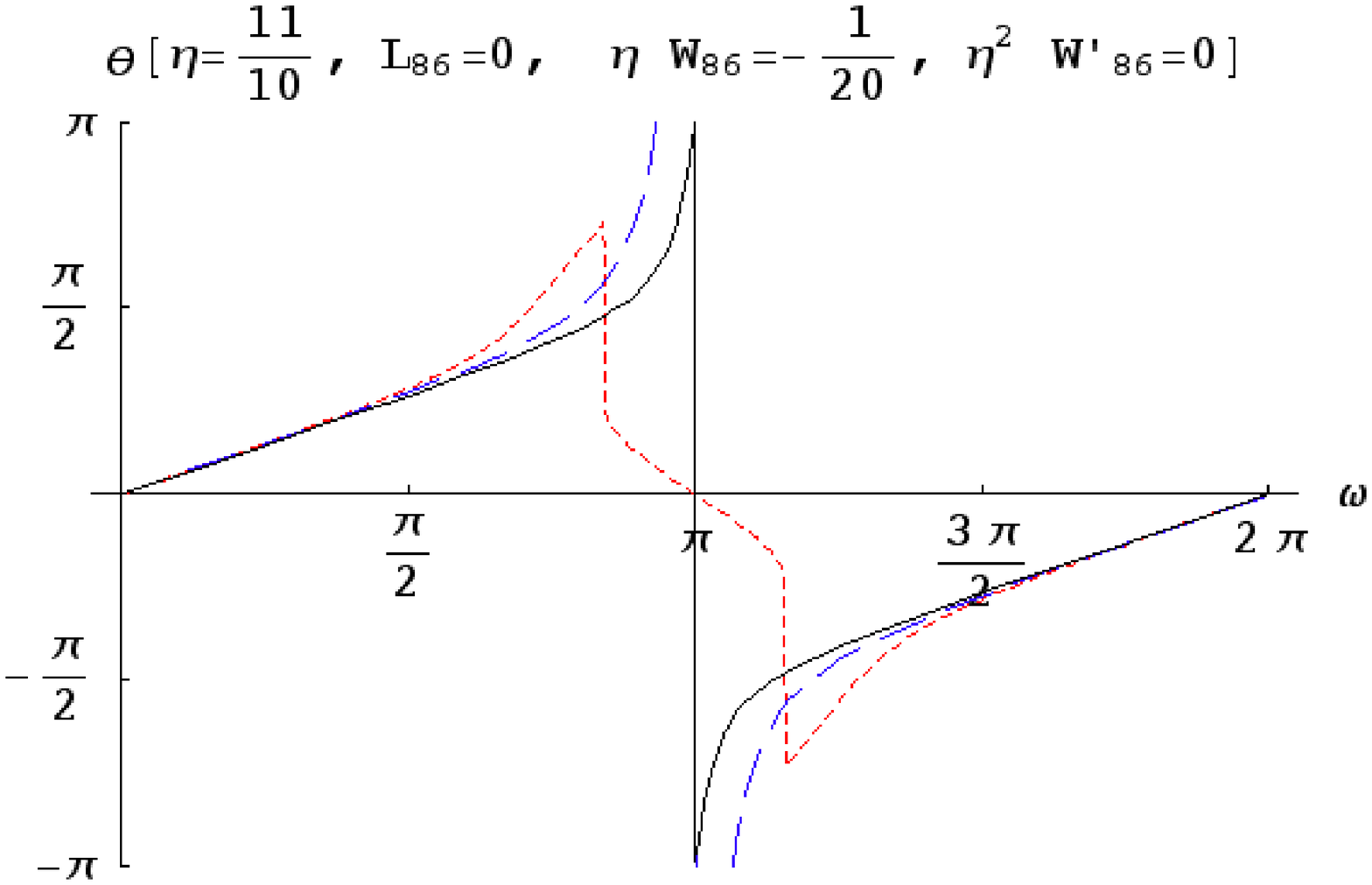}
\includegraphics[height=5cm,width=5cm]{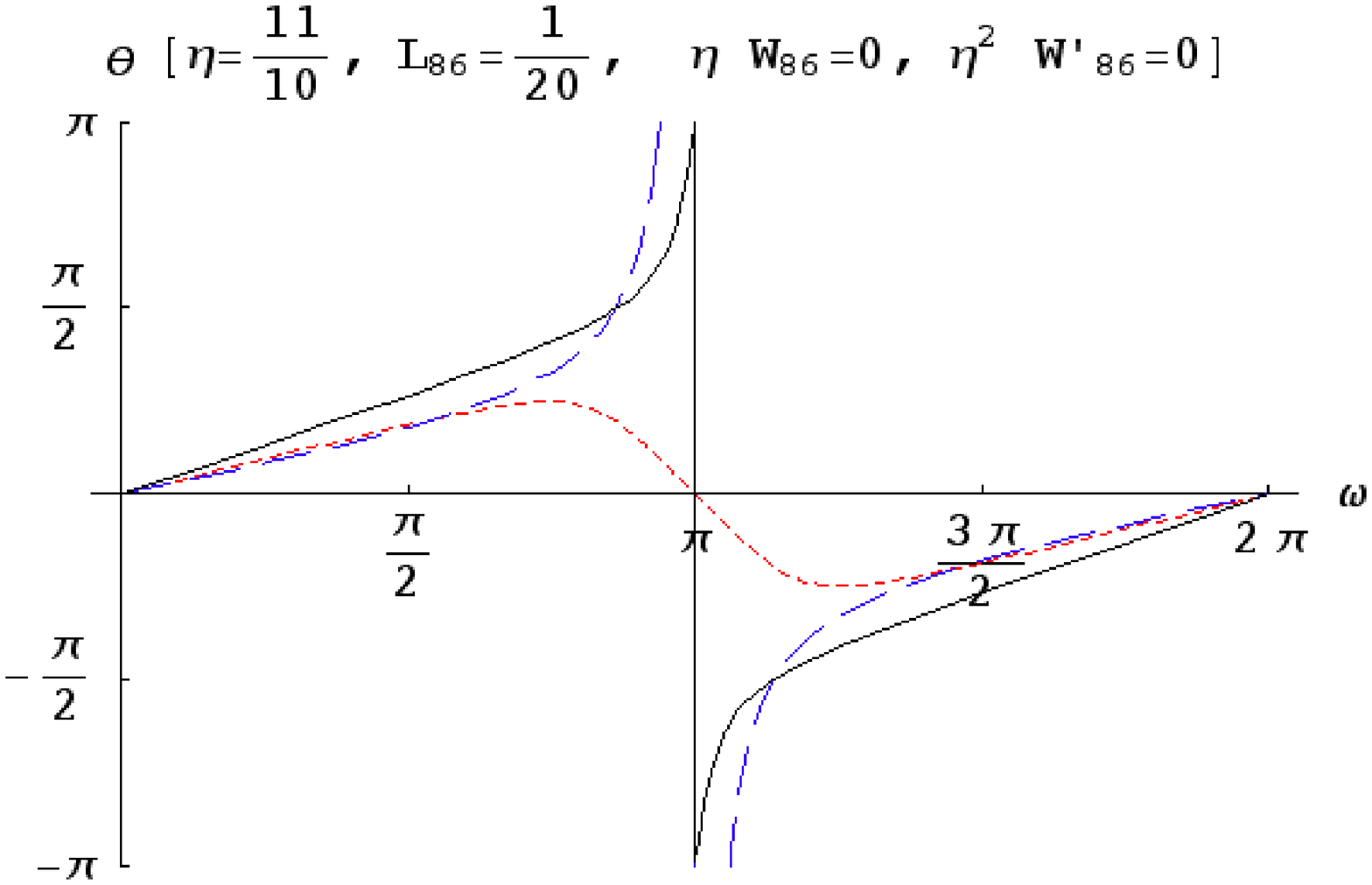}
\includegraphics[height=5cm,width=5cm]{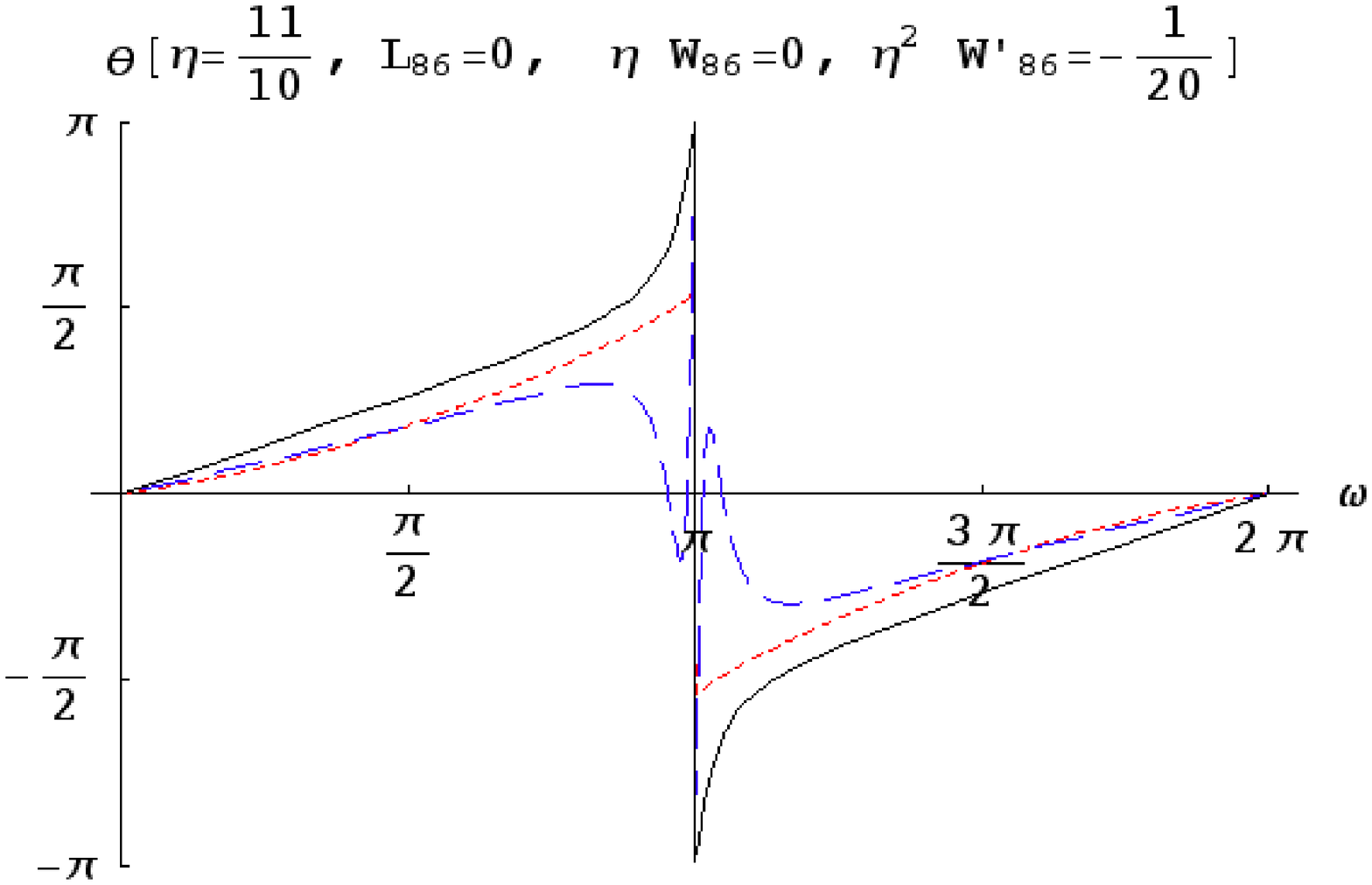}
\includegraphics[height=5cm,width=5cm]{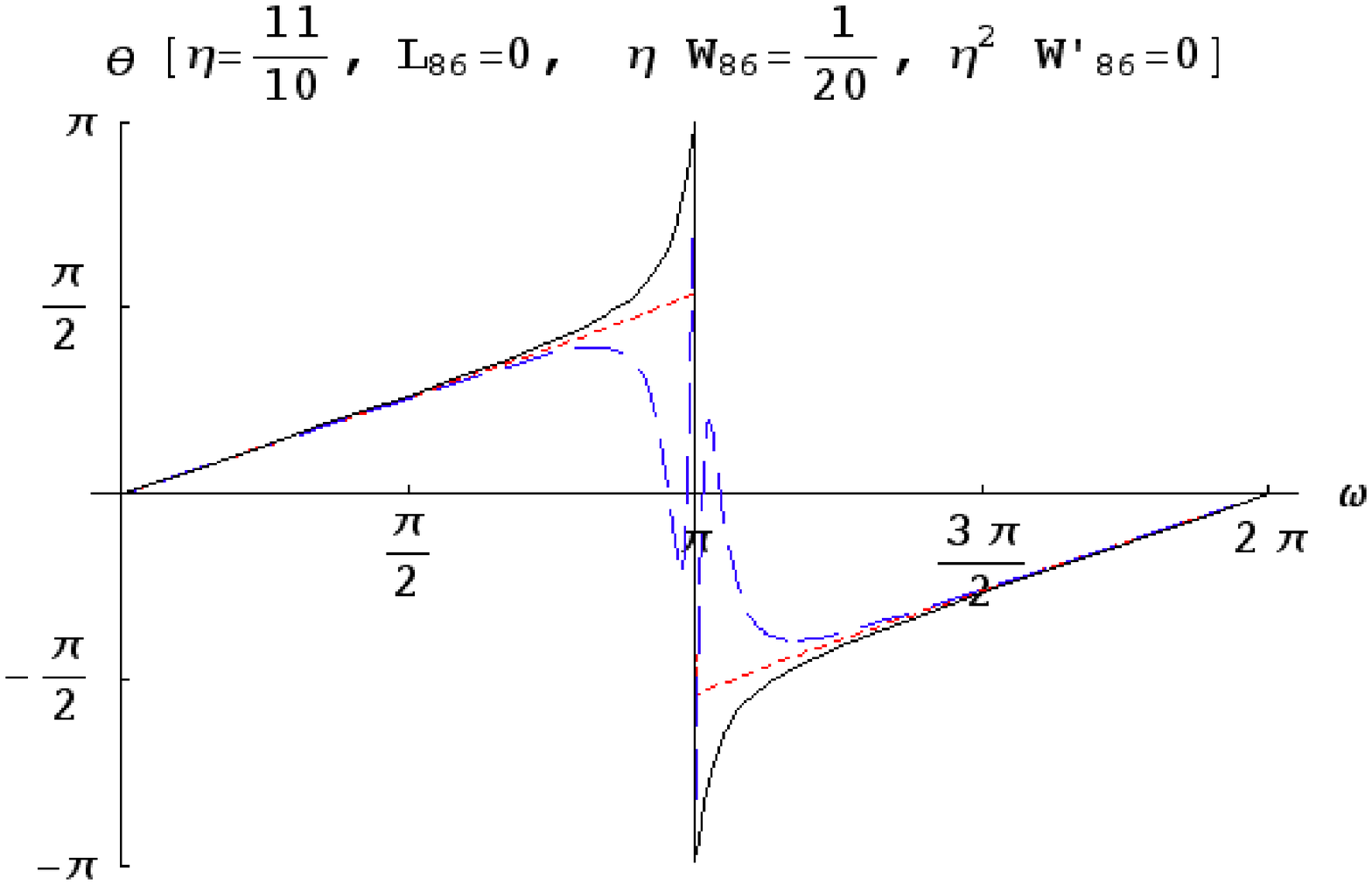}
\includegraphics[height=5cm,width=5cm]{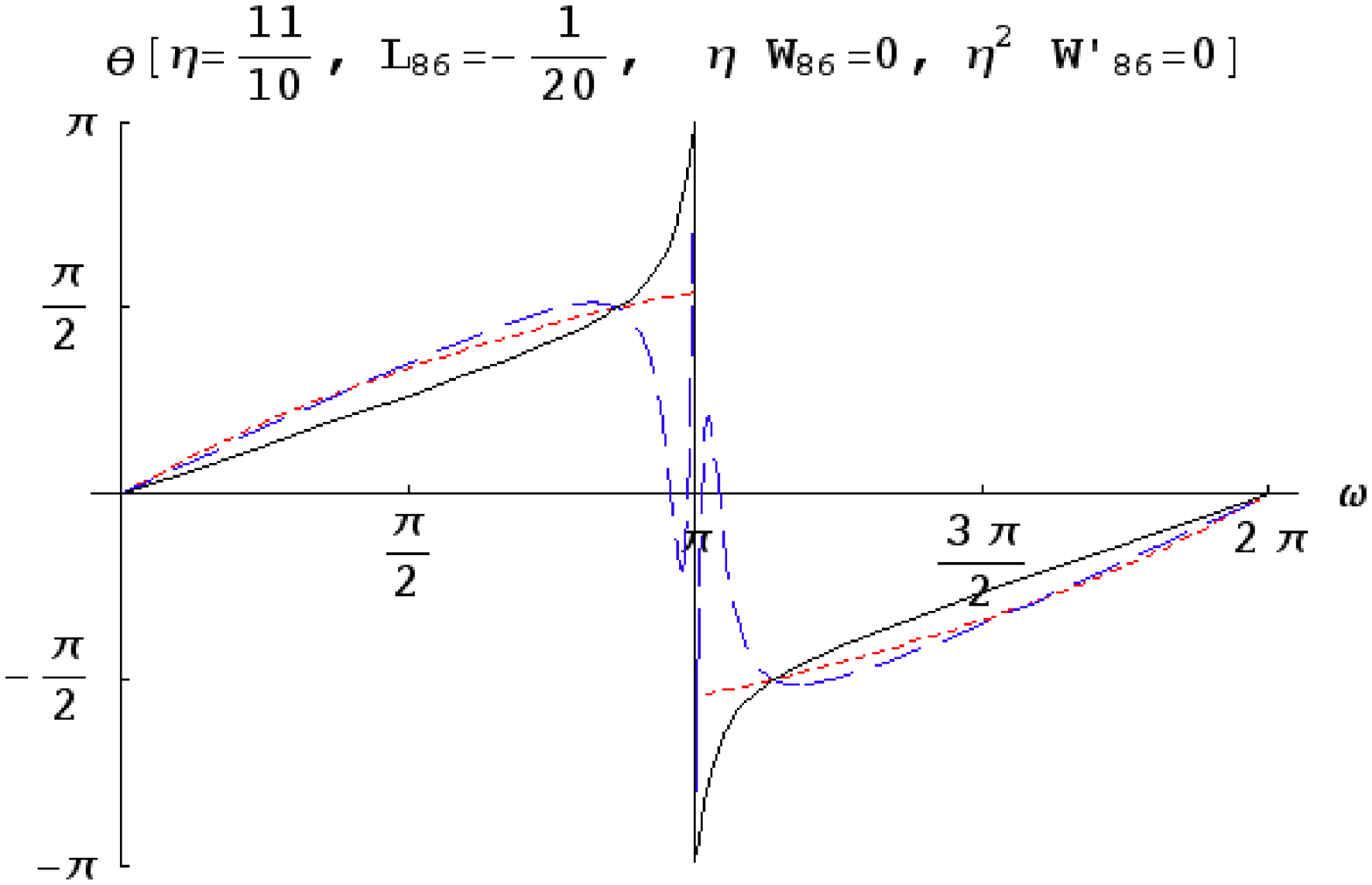}
\caption{\label{fig:NLOvacuum-fetacr} NLO solution for the vacuum orientation $\theta$ 
as a function of $\omega$ for various choices
of the parameters and $\eta\sim 1$. 
Dashed-blue, dotted-red, full-black as before. 
Instabilities are clearly displayed in the NLO analytical solution.
}
\end{figure}

If we switch on $\delta\omega$, the potential does not depend anymore 
only on $\cos(\theta)$, but also on $\sin(\theta)$. This breaks explicitly the symmetry
$\pm \theta$. It is clear that the sign of $\theta$ will be opposite of the sign
of $\delta\omega$ in order to produce in both cases a negative contribution to the potential 
$\delta\omega\sin(\theta)$. We will not try to find the full solution in this case.
In \cite{Munster:2004am} the first correction has been computed.
Here instead we repeat in figure \ref{fig:NLOvacuum-fetacr} a plot analogous to the one 
in figure \ref{fig:NLOvacuum-feta}, but now for $\eta$ very near 1. 
The analytical NLO solution (dashed-blue) display clear instabilities, as expected,
being simply a correction to the LO vacuum. It is not anymore relyable,
when a more complicated structure of minima can develop.
The numerical solution (dotted-red) is very interesting: it shows either a jump at $\omega=\pi$
(see the plots in the second and fourth row in figure \ref{fig:NLOvacuum-fetacr}, here $c_2>0$), 
or a more gradual change (first and third row in figure \ref{fig:NLOvacuum-fetacr}, here $c_2<0$) 
that involves a temporary transition through a stable minima in $\theta=0$. 

\begin{figure}
\includegraphics[height=8cm,width=8cm]{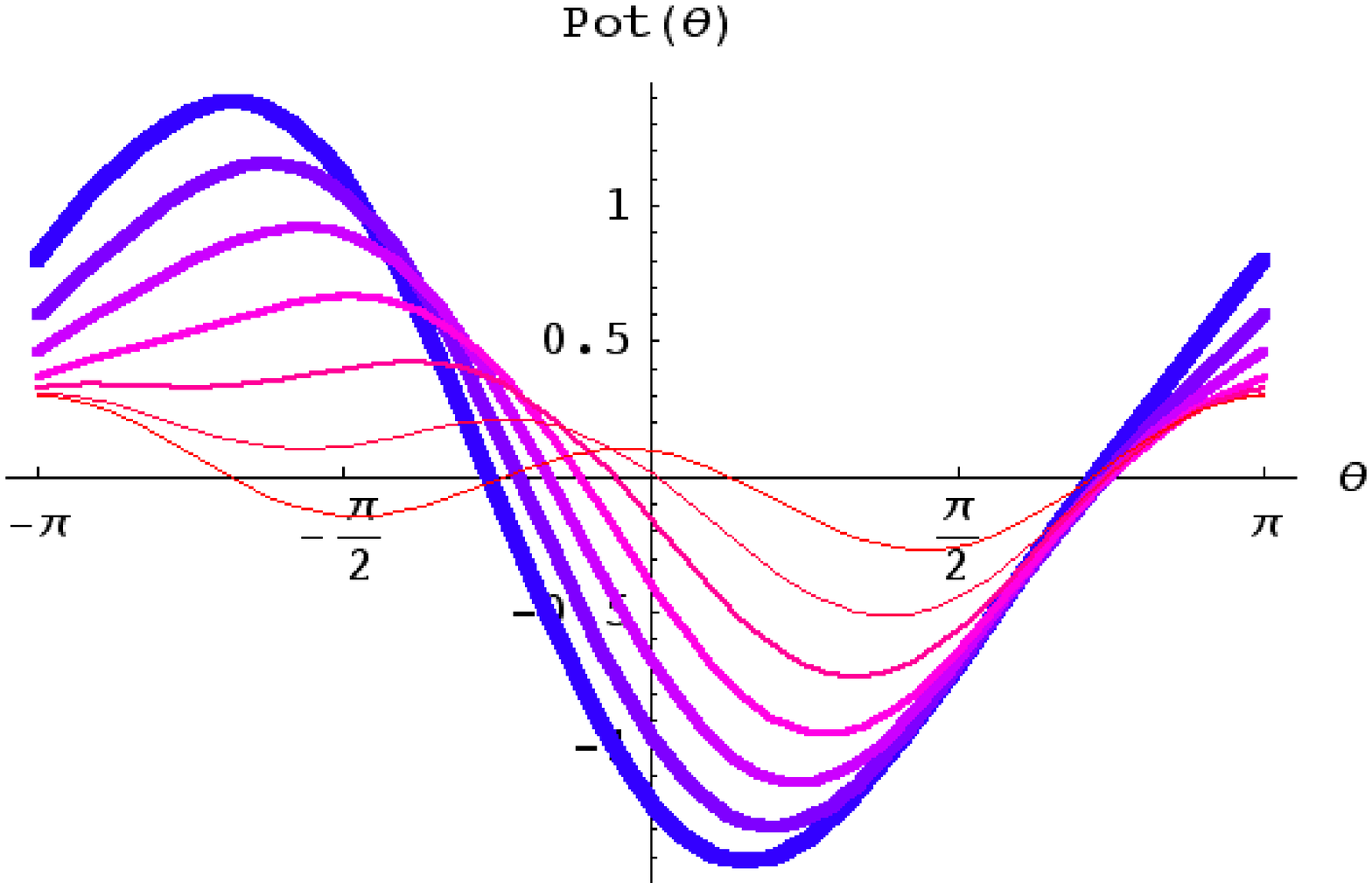}
\caption{\label{fig:pot2min1} Potential ${\cal V}_{\chi}(\theta)$ evolving from $\omega=\pi/2$ 
(thick-blue) to $\omega=\pi$ (thin-red). The other parameters are 
$\eta=9/10$, $\chi\eta^2 W'_{86}=-1/20$, $L_{86}=W_{86}=0$.}
\end{figure}

The first pattern (a sudden jump) corresponds to $c_2>0$. In fact for these 
parameters an Aoki phase is possible, therefore $\theta=0$ is a local maximum, and
the vacuum can only choose between the 2 Aoki vacua at $\theta = \pm \cos^{-1} \epsilon$. Which 
one is chosen, depends on the sign of $\omega-\pi\sim\delta\omega$. 
In figure \ref{fig:pot2min1} we follow what happens in details by looking at the potential
as a function of $\theta$. Going from thick-blue to thin-red,
$\omega$ goes from $\pi/2$ to  $\pi$. The minimum generated by the twisted term
at $\omega=\pi/2$ evolves smoothly into the right Aoki minimum. At $\omega=\pi$ the potential
is the one typical of the Aoki phase, with 2 minima corresponding to
$\theta=\pm \cos^{-1}(\epsilon)$. If $\delta\omega$ changes sign, the vacuum tunnels
between the two minima, and flips the sign of the condensate 
$\langle \bar{\psi} \gamma_5 \tau_3 \psi \rangle$.

\begin{figure}
\includegraphics[height=8cm,width=8cm]{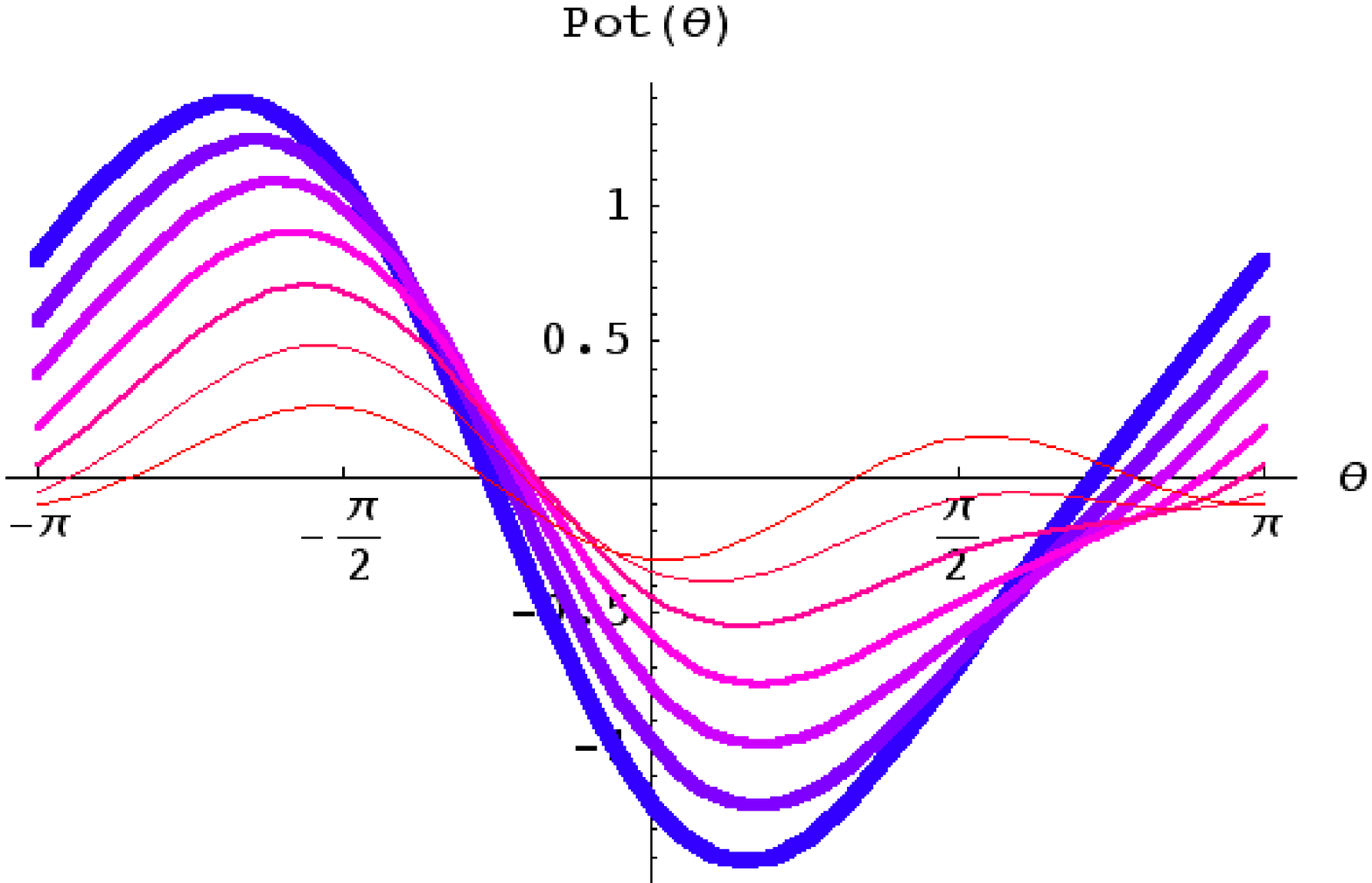}
\caption{\label{fig:pot2min2b}Potential ${\cal V}_{\chi}(\theta)$ evolving from $\omega=\pi/2$ 
(thick-blue) to $\omega=\pi$ (thin-red). The other parameters are 
$\eta=9/10$, $\chi L_{86}=1/20$, $W'_{86}=W_{86}=0$.}
\end{figure}

\begin{figure}
\includegraphics[height=8cm,width=8cm]{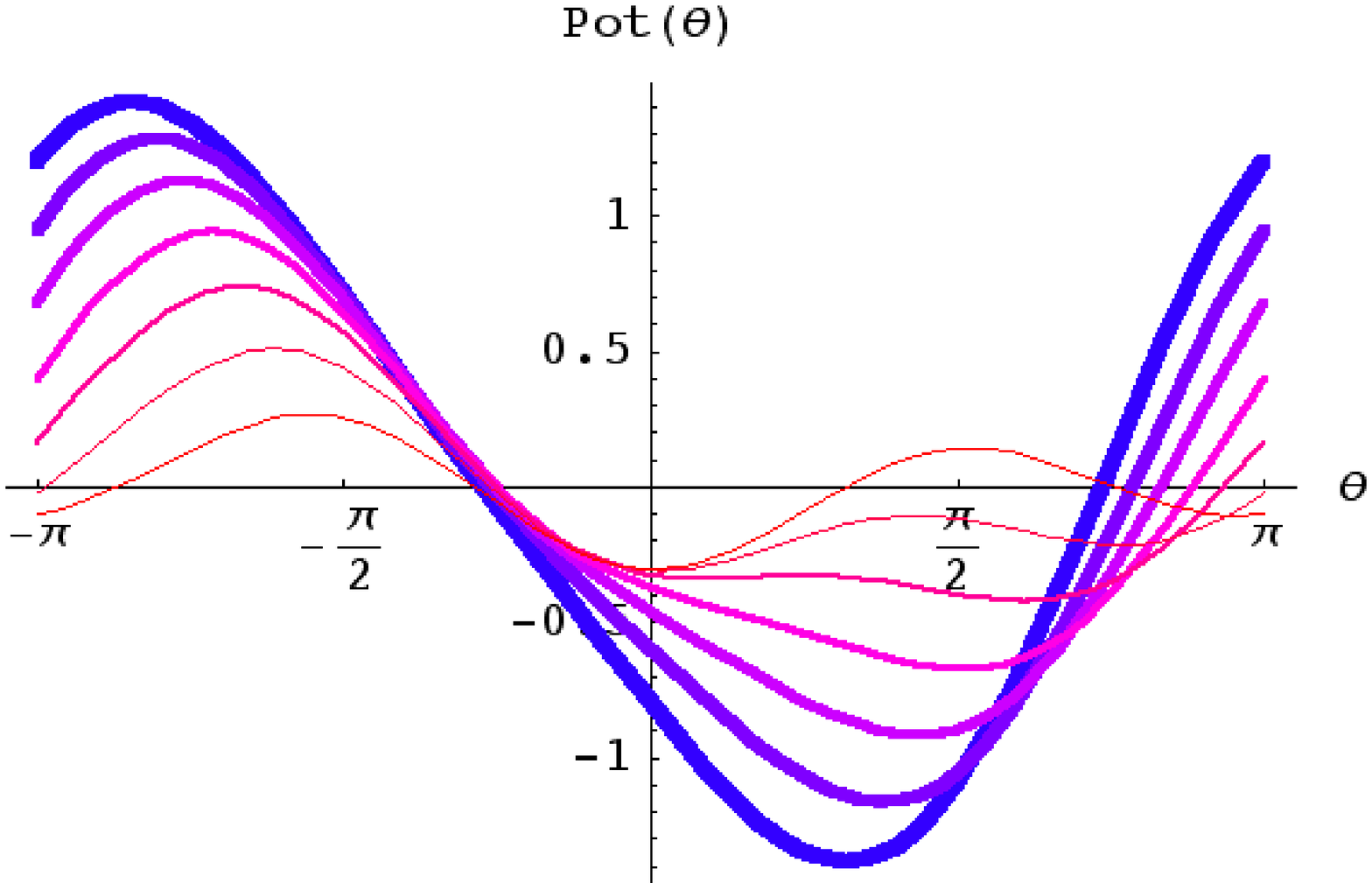}
\caption{\label{fig:pot2min2a}Potential ${\cal V}_{\chi}(\theta)$ evolving from $\omega=\pi/2$ 
(thick-blue) to $\omega=\pi$ (thin-red). The other parameters are 
$\eta=9/10$, $\chi\eta^2 W'_{86}=1/20$, $L_{86}=W_{86}=0$.}
\end{figure}

In the second scenario $c_2<0$ and no Aoki phase is
possible, thus $\theta=0$ is a local minimum at $\omega=\pi$. When the twisted term is
still strong, for instance around $\omega=\pi/2$, it imposes a vacuum far from $\theta=0$.
But as soon as the twisted term is weak enough, the normal vacuum $\theta=0$ becomes again
the true solution. In this scenario the transition can be smooth, but it can also happen
in two separate jumps, signaling a possible coexistence of two local minima for some
finite value of $\pi$. This is a new phenomena, which does not occur in absence of
twisting, and it is worth analyzing more closely. The first possibility is displayed in 
figure \ref{fig:pot2min2b}: the minimum at $\omega=\pi/2$ is essentially the LO minimum dictated
by the twisted term. Approaching $\omega=\pi$ such minimum is smoothly deformed to the trivial
one $\theta=0$. Although another minimum develops for larger $\omega$, it plays no role.
The second picture can be seen in figure \ref{fig:pot2min2a}. Here the unique minimum at
$\omega=\pi/2$ is smoothly deformed to the non trivial minimum near $\theta=\pi$, which is however 
unstable, and the transition to the true one at $\theta=0$ occurs with a jump. 
In this scenario decreasing the twisted term may have the counterintuitive effect of 
producing a larger $\langle \bar{\psi} \gamma_5 \tau_3 \psi \rangle$, although 
only for a short range.
Unfortunately we could not  find a full
characterization in terms of LEC's,  of these last two behaviors. 

Now we should add some comment about the possible values of the parameters that may occur
in practice. According -- for instance -- to simulations performed in 
\cite{Farchioni:2003nf,Farchioni:2004tv}
pion masses between $375$ and $670$ MeV  correspond to 
$\chi$ between $11$ and $36$. In the same reference  also the combination 
$2 L_6+ L_8$ have been estimated\footnote{From data at $\beta=5.1$ and no continuum 
extrapolation has been attempted.} 
to be approximately
$\sim 4.2 (1.1)\cdot 10^{-3}$ (at the renormalization scale $\mu=6 F$). This combination
depends on the renormalization scale, and at present no estimate of any $W's$ is available.
However this justify our choice 
(for illustration purposes) of
plotting the potential in the range $\chi X_{86} = \pm \frac{1}{20}$, ($X=L,W,W'$).

\section{Pion masses and pion mass splitting}
\label{sec:Masses}
In this section we compute the pion masses from the Lagrangian (\ref{eq:ChLag}), including
$O(a^2)$ lattice artifacts in a twisted direction. The expansion is performed around the 
vacuum computed in section \ref{sec:LOvacuum} and \ref{sec:NLOvacuum}.  
As discussed in the previous section, we 
expect this description to be good for any $\eta$ and $\omega$, as long as $\chi$ and 
$\chi\eta$ are small enough (i.e. we are in the ChPT regime), 
and the combination $(\eta^2 + 2 \eta \cos(\omega) + 1)$ is not too small. 
The latter condition means that this expansion cannot describe the ``critical'' region
where both $\delta\omega$ and $\delta\eta$ are $\sim c_2$.
There it is necessary to make sure first that the true vacuum has been recognized
(see section \ref{sec:CritReg}).

With these restrictions in mind, we proceed to the computation of the pion mass. For that
we need only the Lagrangian and we can follow the simple method described for instance
in \cite{Donoghue:1992dd} (sec. $VI-2$). We expand the LO Lagrangian ${\cal L}_{LO}$ 
around the vacuum computed to NLO order,
while for the NLO part of the Lagrangian ${\cal L}_{NLO}$ we only need the vacuum at LO.
The NLO Vacuum -- inserted in ${\cal L}_{LO}$ -- is necessary
to cancel the terms linear in the fields $\pi(x)$, coming from the NLO Lagrangian
${\cal L}_{NLO}$. 
To these two components we must add the effective Lagrangian coming from the loop
integration ${\cal L}_{loop}$. 
The computation of ${\cal L}_{loop}$ is easy, if one observes that the LO Lagrangian 
expanded around the vacuum at LO is equivalent to the untwisted Chiral Lagrangian at LO,
without Wilson term, but with a mass $m^*$ such that:
\[
\frac{2 B_0 m^*}{F^2}= 
\chi 
\sqrt{{{\eta}^2}+2  \cos(\omega )  \eta +1} \equiv \chi^*
\]
This is the new version of the ``rule'' ($\chi \rightarrow \chi + \rho$) 
suggested in \cite{Rupak:2002sm} to compute the logarithmic term in WChPT from ordinary ChPT.
We have then: 
\[
{\cal L}_{loop} = 
\frac{1}{2} \partial_{\mu} \vec{\pi} \cdot \partial^{\mu} \vec{\pi} 
(-\frac{2 B_0 m^*}{24 \pi^2 F^2} \Delta(m^*))
-
\frac{1}{2} \vec{\pi}^2 (-\frac{2 (B_0 m^*)^2 }{24 \pi^2 F^2} \Delta(m^*)),
\]
where, in dimensional regularization, we defined,
\[
\Delta(2 B_0 m^*)= 
\log(\frac{2   B_0   m^*}{\Lambda^2 })+\frac{2}{d-4 }-\log(4   \pi )+\gamma_E -1.
\]
The pion wave function renormalization -- which is defined by
$\pi_R(x) = Z_{\pi}^{-1/2}\pi(x)$ -- results:
\begin{eqnarray*}
Z_{\pi} &=& 1 - 
\chi \frac{8  ((\eta   \cos(\omega )+1)  {L_{45}}+\eta   (\eta +\cos(\omega ))  {W_{45}}) }
{{\sqrt{{{\eta}^2}+2  \cos(\omega )  \eta +1}}}\\
&&+ 
\chi 
\frac{{\sqrt{{{\eta }^2}+2  \cos(\omega )  \eta +1}} }{24  {{\pi }^2}} \Delta(2 B_0 m^*)
+{O}({{\chi }^2}),
\end{eqnarray*}
which converges to the result in \cite{Munster:2003ba} in the limit $\eta<<1$. No new LEC
is involved here. If we call $\pi^{\pm}$ the pion associated to the untwisted directions
$\tau_{1,2}$ we find:
\begin{eqnarray} \label{eq:pionmass}
\frac{m^2_{\pi^{\pm}}}{F^2} &=& 
\chi {\sqrt{{{\eta }^2}+2  \cos(\omega )  \eta +1}}-
\chi^2 [
\frac{1}{{{\eta }^2}+2  \cos(\omega )  \eta +1} \\
&& (8  (
-2  {{\bar L}_{86}}  {{(\eta   \cos(\omega )+1)}^2}
+
(\cos(2 \omega )  {{\eta }^2}+2  {{\eta }^2}+({{\eta }^2}+3)  \cos(\omega )  \eta +1)  {{\bar L}_{45}}
+ \nonumber\\
&&+
\eta   (\eta +\cos(\omega ))  
(({{\eta }^2}+2  \cos(\omega )  \eta +1)  {{\bar W}_{45}}
-
2  (\eta  
\cos(\omega )+1)  {{\bar W}_{86}}-2  \eta   (\eta +\cos(\omega ))  {{\bar W'}_{86}})))
]+ \nonumber\\
&&+ 
\chi^2 [
\frac{({{\eta }^2}+2  \cos(\omega )  \eta +1)}{32  {{\pi }^2}}
]\log\Big(  \chi \sqrt{{{\eta}^2}+2  \cos(\omega )  \eta +1}  \Big)
+{O}({{\chi }^3}). \nonumber
\end{eqnarray}
Once again in the limit $\eta\rightarrow 0$ the result in \cite{Munster:2003ba} is recovered.
Here we introduced the renormalized LEC at the scale $F^2$ as defined in \cite{Farchioni:2003bx}:
\begin{eqnarray*}
{\bar L}_{45} &=& L_{45} - (C_5 + 2 C_4) \Delta(F^2) \\
{\bar W}_{45} &=& W_{45} - (D_5 + 2 D_4) \Delta(F^2) \\
{\bar L}_{86} &=& L_{86} - (C_8 + 2 C_6) \Delta(F^2) \\
{\bar W}_{86} &=& W_{86} - (D_8 + 2 D_6) \Delta(F^2) \\
{\bar W'}_{86} &=& W'_{86} - (C_8 + 2 C_6) \Delta(F^2) \\
C_4 &=& \frac{1}{256 \pi^2}, \;\;\;\; 
C_5 = \frac{1}{128 \pi^2}, \;\;\;\; 
C_6 = \frac{3}{1024 \pi^2}, \;\;\;\; 
C_8=0 \;\;\;\; D_i = 2 C_i.
\end{eqnarray*}
Notice that the LEC $W'_i$ are renormalized with the same coefficients $C_i$ as
the $L_i$. Notice also that, although the functional forms (in $\eta$ and $\omega$) in front of the
various LEC's are very different, the coefficients $C_i$ and $D_i$ are such that
they sum up to precisely the functional form in front of the loop term. 
This is a non trivial check of consistency for the calculation.

The most interesting aspect of the formula for the pion mass (\ref{eq:pionmass})
is that each of the LEC's enters the expression with a different functional 
dependence on $\eta$ and $\omega$. These are simulation parameters that can be 
{\em in principle} freely changed in Monte Carlo simulations. This allows {\em in principle}
a separate determination of each of the LEC. This somehow reminds the successful idea
of selecting special combinations of LEC from Partially Quenched simulations \cite{Sharpe:2000bc}. 
But the situation is very different here, and
whether this could be really exploited in numerical simulations, cannot be answered here.

A twisted Wilson term breaks explicitly the flavor symmetry. Thus one expects a splitting
in the degeneracy of the pion masses. 
If we call $\pi^{0}$ the pion associated to the twisted directions
$\tau_{3}$ we find:
\begin{eqnarray} \label{eq:massplit}
\frac{m^2_{\pi^{\pm}}-m^2_{\pi^0}}{F^2} &=&
\frac{16 {{\eta }^2}  {{\chi }^2}  {{\sin}^2}(\omega )  
({{\bar L}_{86}}-{{\bar W}_{86}}+{{\bar W'}_{86}})}{{{\eta }^2}+2 
\cos(\omega )  \eta +1} \\
&\overrightarrow{\eta \rightarrow 0}&
\;\;\;\;\;\;\;\;\;\;
64 a^2 \sin(\omega)^2 \frac{W_0^2 ({{\bar L}_{86}}-{{\bar W}_{86}}+{{\bar W'}_{86}})}{F^4}.
\nonumber
\end{eqnarray}
The most interesting aspect of this formula is that the pion mass splitting is completely
determined (to this order) 
by precisely the same combination of LEC's that decides about the possibility 
of an Aoki phase. In particular $m^2_{\pi^{\pm}}>m^2_{\pi^0}$ predicts $c_2<0$ and no Aoki phase.
Recent unquenched simulations \cite{Farchioni:2004us,Ilgenfritz:2003gw} suggest that
there is indeed no Aoki phase at any interesting $\beta$. It would be nice to
confirm this picture through the measurement of the pion masses. Unfortunately  $m_{\pi^0}$ is 
a bit more difficult to measure than the other two in tmQCD, since it may have
stronger contaminations from heavier states, but it should not be impossible.

As already mentioned, the determinations of $2 L_6 + L_8$, which are available both
from the lattice \cite{Farchioni:2004tv} and from experiments
\cite{Bijnens:1994qh}, are not enough for our purposes, because such combination
alone depends on the renormalization scale, while the combination 
in (\ref{eq:massplit}) does not.

\subsection{Pion masses near the critical region}
Our final task is to try to make connection between the pion masses computed in
(\ref{eq:pionmass},\ref{eq:massplit}) and the computations in \cite{Sharpe:1998xm}.
As already said, we do not expect this to succeed, because the vacuum around 
which we expand is blind to the appearance of a complicated structure of minima.
However some partial result can be obtained.

We employ again the expansion (\ref{eq:paramcrit}) that we used before for the critical region.
In these variables the ``charged'' pion masses become
\begin{equation} \label{eq:mpi12crit}
\frac{m^2_{\pi^{\pm}}}{F^2} =
{{\chi }^2}  
\bigg(\frac{16  ({L_{86}}-{W_{86}}+{Z_{86}})  {{{\delta \eta }}^2}}
{{{{\delta \eta }}^2}+{{{\delta\omega }}^2}}+
{\sqrt{{{{\delta \eta }}^2}+{{{\delta \omega }}^2}}}\bigg),
\end{equation}
while the ``neutral'' pion is:
\begin{equation} \label{eq:mpi3crit}
\frac{m^2_{\pi^{0}}}{F^2} =
{{\chi }^2}  
\bigg(\frac{16  ({L_{86}}-{W_{86}}+{Z_{86}})
({\delta \eta}^2-{\delta \omega}^2) }
{{{{\delta \eta }}^2}+{{{\delta \omega }}^2}}+
{\sqrt{{{{\delta \eta }}^2}+{{{\delta\omega }}^2}}}\bigg).
\end{equation}
If we recall the identification (\ref{eq:ident}) and set $\delta\omega\rightarrow 0$
we have three degenerate pions which reproduce the masses computed
in \cite{Sharpe:1998xm} when the vacuum is trivial. More precisely
we reproduce the formula (4.18) of \cite{Sharpe:1998xm}, when $c_2<0$ (i.e. no Aoki phase)
and the formula (4.14b) of \cite{Sharpe:1998xm}, when $c_2>0$ and $|\epsilon |>1$
(i.e. Aoki phase is possible, but not yet reached).
When $c_2>0$ and $|\epsilon |<1$, our formulas 
(\ref{eq:mpi12crit},\ref{eq:mpi3crit}) do not make sense, because they
predict negative masses.

However we can do better. We can approach the critical point trying to
follow the vacuum (\ref{eq:solOmegaNLO}) along a path that
merges into the Aoki vacuum. This is obtained 
when $\eta\rightarrow 1$ faster than $\omega\rightarrow\pi$.
Therefore, instead of (\ref{eq:paramcrit}), we use the parameterization:
\begin{equation} \label{eq:paramcrit2}
\eta = 1 + \chi^2 \delta\eta, 
\;\;\;\;\;\;\;\;\;\; 
\omega = \pi + \chi \delta\omega. 
\end{equation}
In this way we get for the pion masses:
\begin{equation} \label{eq:mpi123crit2}
\frac{m^2_{\pi^{\pm}}}{F^2} =
{{\chi }^2}  
{\sqrt{{{{\delta \omega }}^2}}},  \;\;\;\;\;\;\;\;\;\;
\frac{m^2_{\pi^{0}}}{F^2} =
{{\chi }^2}  
\Big({\sqrt{{{{\delta \omega }}^2}}}-16  ({L_{86}}-{W_{86}}+{Z_{86}})\Big).
\end{equation}
Now, setting $\delta\omega\rightarrow 0$, 
we find the expected massles charged pions. 
Instead, the neutral pion is not as expected. The value of $m_{\pi^0}$
in (\ref{eq:mpi123crit2}) agrees with (4.14a) in \cite{Sharpe:1998xm}
only for vanishing $\epsilon$. 
Also missing is a better understanding of the transition described in figure
\ref{fig:pot2min2a}, over which we have no analytical control.
On the whole, however, the formula (\ref{eq:pionmass}) provides more informations
about the critical region, than one could expect.

\section{Conclusions}
\label{sec:disc}
We have proposed a method to organize a ChPT calculation which is convenient in the case
of tmQCD and has a better range of applicability than the ordinary expansion around the 
trivial vacuum. We have computed the vacuum orientation to NLO and compared with
numerical results which give a rough estimate of the applicability of the expansion. We have
checked that -- within the range of validity of a ChPT expansion -- 
no other orientations are possible for the vacuum,
apart from combinations of $1_2$ and $\tau_3$. 
As long as we are not too close to the point $\eta=1$ and $\omega=\pi$ 
(i.e. where $B_0 m$ and $W_0 a$ are equal in size and point in the same direction), 
the potential has a single minimum,
which is essentially the LO vacuum plus corrections. When we approach the critical region,
a more complicated structure of minima is possible, and any perturbative description 
breaks down. We show -- mainly numerically -- as the LO vacuum is smoothly connected
to an Aoki phase, if present. If no Aoki phase is possible, we displayed two different
scenarios (one smooth, and one involving a transition) 
in which the LO vacuum at non zero twist is connected to the trivial vacuum. 

In the second part we used the vacuum computed before to determine the pion masses. 
We noticed that the introduction of a new degree of freedom allows in principle a better 
determination of the LEC's. Having performed the computation to $O(a^2)$, we could also determine
the splitting between the pion masses in the twisted ($m_{\pi^0}$) 
and in the untwisted ($m_{\pi^{\pm}}$) directions. 
Surprisingly, this splitting -- which should be measurable even for quite heavy pion masses --
gives direct information about the phase structure of the system in the deep unphysical
region dominated by lattice artifacts.
Finally we tryed to extrapolate our formulas for the pion masses inside the
critical region, obtaining some partial results.

\section*{Acknowledgments}
I thank Karl Jansen, Istvan Montvay, Gernot M\"unster, Enno Scholz, Andrea Shindler and 
all the participants to the ``Discussion Seminars'' organized by Ulli Wolff at HU 
for stimulating discussions. 
I am also grateful to Noam Shoresh for communications. 
I acknowledge support from DFG through the Sonderforschungsbereich
``Computational Particle Physics'' (SFB/TR 9).
\bibliography{tmWa2chpt}

\end{document}